\documentclass[11pt]{article}

\usepackage{geometry}
\usepackage[toc,page]{appendix}

\usepackage{fancyhdr}
\usepackage{verbatim}

\usepackage{bbm}
\usepackage{amssymb}
\usepackage{amsmath}
\usepackage{amsthm}
\usepackage{dsfont}
\usepackage[dvipsnames]{xcolor}
\usepackage{ stmaryrd }
\usepackage{ bbold }

\usepackage{titlesec}
\titlelabel{\thetitle.\quad} %in order to have section 1. rather than 1

\usepackage{multibib}
\newcites{supp}{References}

\usepackage{graphicx}

\usepackage{mathtools}
\DeclarePairedDelimiter\floor{\lfloor}{\rfloor}

\usepackage{tikz}
\usepackage{pgfplotstable}
\usepackage{pgfplots}
\usepackage[nomessages]{fp}% http://ctan.org/pkg/fp
\usepackage{ifthen}
\usepackage{standalone}

\usetikzlibrary{external}
\tikzexternalenable

\usepackage{algorithm,algpseudocode}

\usepackage[colorlinks = true, linkcolor = blue,
            urlcolor  = blue,
            citecolor = blue,
            anchorcolor = blue,
            pdfborder={0 0 0},
             plainpages=false,pdfpagelabels]{hyperref}

\usepackage[round]{natbib}
\usepackage{url}

\makeatletter
\newcommand{\oset}[3][0ex]{%
  \mathrel{\mathop{#3}\limits^{
    \vbox to#1{\kern-2\ex@
    \hbox{$\scriptstyle#2$}\vss}}}}
\makeatother

\newtheorem{defn}{Definition}

\newtheorem{propr}{Proposition}

\definecolor{corrections}{RGB}{0,0,0}  %{0,0,0}{178,34,34}

\begin{document}

{
 \title{\textbf{Changepoint Detection on a Graph of Time Series}} 
 \author{Karl L. Hallgren$^{1}$\thanks{Email: karl.hallgren17@imperial.ac.uk}, Nicholas A. Heard$^1$ and Melissa J. Turcotte$^2$ \vspace{.5cm} \\
    $^1$Department of Mathematics, Imperial College London \\
    $^2$Microsoft Corporation %, Redmond, Washington, United States
}
\date{\vspace{-5ex}}
\maketitle
}

%%%%%%%%%%

\begin{abstract}
When analysing multiple time series that may be subject to changepoints, it is sometimes possible to specify \textit{a priori}, by means of a graph, which pairs of time series are likely to be impacted by simultaneous changepoints. 
This article proposes an informative prior for changepoints which encodes the information contained in the graph, inducing a changepoint model for multiple time series that borrows strength across clusters of connected time series  
to detect weak signals for synchronous changepoints. The graphical model for changepoints is further extended to allow dependence between nearby but not necessarily synchronous changepoints across neighbouring time series in the graph.  
A  novel reversible jump Markov chain Monte Carlo (MCMC) algorithm making use of auxiliary variables is proposed to sample from the graphical changepoint model. 
The merit of the proposed approach is demonstrated through a changepoint analysis of computer network authentication logs from Los Alamos National Laboratory (LANL), demonstrating an improvement at detecting weak signals for network intrusions  across users linked by network connectivity, whilst limiting the number of false alerts. 
\end{abstract}


\section{Introduction}

Consider $N>1$ time series of random observations
\begin{align}
\label{eq:introdata}
\{ x_{i, t} \, | \,  1 \leqslant i \leqslant N,  t \geqslant 0\}
\end{align}
which are subject to changepoints. This article will suppose the existence of an underlying graph $G$ on $N$ nodes corresponding to each of the time series, such that changepoints are % \textit{a priori} 
believed to occur simultaneously or closely together in time for time series connected by edges in $G$.

A motivating application for considering such dependencies is the task of changepoint detection in cyber-security. To identify the presence of a network intrusion, it is informative to monitor for changes in the authentication activity of each user in the network.
However, cyber data often exhibit much variability 
and apparent changes are not guaranteed to correspond to an attack. As a result, to limit the number of false alerts and yet not overlook weak signals from genuine, small  attack footprints, it is key to incorporate expert knowledge in the change detection procedure. A commonly held belief of security experts is that attacks are \textit{a priori} likely to be identified through quasi-simultaneous changes in the behaviour of users that are linked by network connectivity \citep{Sexton2015AttackCD}. Hence, it is of interest to encode a changepoint prior by means of a graph $G$ representing the network of users, such that pairs of connected users in $G$ are \textit{a priori} more likely to be affected by quasi-simultaneous behavioural changes. %, thereby inducing a changepoint model  
%for the authentication data
%that borrows strength across clusters of connected users in $G$ to detect signals for quasi-synchronous changes. 

Limited attention has previously been given to encoding prior beliefs on graph-based dependence structure of discrete-time changepoints across multiple time series. Existing changepoint model for multiple time series, which admit changepoints may simultaneously affect a subset of the time series, typically assume \textit{a priori} changepoint locations are exchangeable across time series \citep{Jeng2012, fearnheadAbnormalSegments, Bolton2018, samworth, detectpanel, Grundy2020HighdimensionalCD}. 
Moreover, with the exception of \citet{fisch}, dependent changepoints across time series are often assumed to perfectly align, which is a limiting assumption in cyber-security monitoring where attacks may span a substantial period of time. %multiple time units. 

More generally, graphical models provide a useful framework for characterising joint distributions for random variables: the nodes of the graph identify the random variables and the edges characterise dependencies among these variables \citep{lauritzen1996}.  
%Areas of application are diverse and include real-life network modelling where edges are natural representations of network connections. 
In particular, graphical models have been employed to encode prior beliefs, for example, in the context of Bayesian variable selection for regression models.  \citet{fanzhang} assumes that covariates lie on an undirected graph and formulates an Ising model prior on the covariate space to incorporate structural information.

This article proposes an informative, graphical model-based prior for changepoints that encodes beliefs on the dependence structure of changepoints across time series (\ref{eq:introdata}). 
%Time series indices are assumed to lie in an undirected graph $G$ that describes which pairs of time series are \textit{a priori} likely to be simultaneous affected by changepoints. 
For practical purposes, changepoints are represented in discrete time by a binary matrix $\boldsymbol{S} = (S_{i, t})$, such that $S_{i, t}$ indicates whether the time point $t$ is a changepoint for the time series with index $i$. Then, extending the standard memoryless prior for changepoints \citep{exactcpt}, independent and identical Markov random fields \citep{lauritzen1996} with respect to $G$ are assumed \textit{a priori} for the columns of $\boldsymbol{S}$. As a result, the model assumes that clusters of time series (according to $G$) are likely to be simultaneously affected by changepoints. Conditional on changepoints, the time series data are assumed to be independent of $G$ and to follow a standard parameteric changepoint model \citep{exactcpt}. A key consequence of the graphical model is that stronger evidence from data is required to infer scattered synchronous changepoints than synchronous changepoints clustered according to $G$. Furthermore, a more general model is proposed that admits related changepoints not occuring at exactly the same time; % to form a priori likely structures across time series; 
the extended model supposes that changepoints may cluster according to $G$ within some finite time windows of possibly unknown lengths, which are specific to each series.

A common approach to sampling changepoints for a single time series is that of \citet{Green1995},   using a reversible jump MCMC algorithm to explore the state space of changepoints: at each iteration of the algorithm, a new changepoint is proposed, or else an existing changepoint is either deleted or shifted to a new position. 
% Alternative approaches to sample changepoints of a single time series include \citet{CHIB1998} and \citet{exactcpt}.
Specifying a joint model for dependent changepoints across multiple time series introduces additional computational challenges that are not present when changepoints are inferred for each time series independently. 
A simulation study will demonstrate that 
it can be impractical to simply propose updates to the changepoints of a randomly chosen time series via one of the moves of \citet{Green1995}. 
To efficiently explore the state space of dependent changepoints, it is necessary to consider joint proposals for changepoints across multiple time series.

We propose an MCMC algorithm making use of auxiliary variables \citep{besagauxi}  to sample from the posterior distribution.  
\citet{Swenden} and \citet{higdon} provide notable examples of use of auxiliary variables in MCMC schemes that improve mixing and convergence for undirected graphical models. In brief, our sampling strategy is the following. The changepoint parameter space is augmented with auxiliary variables that induce clusters of time series  according to the dependence graph $G$. Then, the MCMC algorithm of \citet{Green1995} is extended to sample from the augmented parameter space, such that, at each iteration of the algorithm, a new cluster of changepoints may be proposed or an existing cluster of changepoints may be deleted or shifted. 

Bayesian inference for changepoints quantifies uncertainty about the number and the positions of changepoints. However, in some applications such as cyber-security, it will also be necessary to report a point estimate for changepoint parameters. Yet no existing loss function in the literature seems suitable for taking into account both the number and the positions of changepoints.
To address this gap, we propose using matchings in graphs \citep{Bondy1976} to define a novel loss function for changepoints, which can be used to obtain a point estimate from a posterior sample of candidate changepoints.

The practical benefits of the proposed graphical model are demonstrated via a changepoint analysis of real computer network authentication data from Los Alamos National Laboratory (LANL), where a subset of the data relating to a `red team' exercise provide a proxy for intruder behaviour \citep{akent-2015-enterprise-data}. 
The challenge consists of monitoring for temporal changes in the authentication activity of network users to detect the presence of red team actors. The proposed changepoint prior %graphical changepoint model 
is used to encode beliefs that signals for network intrusions are \textit{a priori} likely to occur at nearby times for users historically linked by previous network connectivity. We show that, as a consequence, the proposed model can detect weak signals for red team activity in the network, whilst limiting the number of false alerts, in contrast with a standard model assuming independence of behavioural changes across users.

Finally, it should be noted that, in contrast with recent changepoint detection methods \citep{ChenZhang, chen2019a, chen2019b, ChenChu}, the focus of this article is not the temporal evolution of a graph subject to changepoints. The graph $G$ represents prior information that can be exploited to detect changepoints in time series. % in multiple time series. 

The remainder of the article is organised as follows. 
Section \ref{sec:motivationalsection} motivates our work with a cyber-security application. %the informative graph-based prior for changepoints introduced in this article. 
Section \ref{sec:changepointanalysisoverview} presents Bayesian changepoint modelling for multiple time series. Section \ref{sec:newprior} introduces a novel,  graph-based informative prior for changepoints. %, which induces a graphical changepoint model for multiple time series. 
Section \ref{sec:inferenceMCMC} proposes an auxiliary variable MCMC sampling strategy. 
Section \ref{sec:bayesestimate} proposes a novel loss function for assessing changepoints. 
%Section \ref{sec:simustudy} presents a simulation study to demonstrate the advantages of the approach. 
Section \ref{sec:cyberapplication} presents results of a changepoint analysis of network authentication data, illustrating the practical benefits of the proposed model. 
%\textcolor{corrections}{Section \ref{sec:simustudy} in the supplementary material \citep{SuppCPGTS} presents a simulation study to demonstrate the advantages of the approach.}
%
%The supplementary material \citep{SuppCPGTS} presents some technical material in 
Appendices \ref{sec:appendixcyber}, \ref{sec:appendixproof} and  \ref{sec:appendixMCMCalgo} present some technical material, and a simulation study in Appendix \ref{sec:appendixsimustudy} demonstrates the model introduced in Section \ref{sec:newprior}.

\section{Motivational application: changepoint detection in cyber-security}
%%%
\label{sec:motivationalsection}

To motivate an informative graph-based prior for changepoints, %which induce a graphical changepoint model, 
we consider an application of changepoint detection in cyber-security. A cyber-attack typically changes the behaviour of connected endpoints on the target computer network \citep{Sexton2015AttackCD}. 
Therefore, to detect the presence of a network intrusion, it is informative to monitor for synchronous, or quasi-synchronous, changes in the behaviour of entities that are \textit{a priori} known to be linked by network connectivity. %connected on the network.

\vspace{-2mm}
%\subsection{Enterprise network data}
\subsection{Change detection in the authentication activity of users}

\citet{akent-2015-enterprise-data} presents a comprehensive data set summarising $58$ days of traffic on the enterprise computer network of Los Alamos National Laboratory (LANL), which is available online at %\url{https://csr.lanl.gov/data/cyber1}.  
\url{https://lanl.ma.ic.ac.uk/data/cyber1}. 
The network authentication data consist of records describing authentication activity of users connecting from one computer to another. The occurrence of a `red team' penetration testing operation during the data collection period makes these data suitable for testing network intrusion detection methods. Further details on the data are given in Appendix \ref{sec:appendixcyberdata}. % in the supplementary material \citep{SuppCPGTS}. 
%Appendix \ref{sec:appendixcyber} \citep{SuppCPGTS}. 
%

Let $V$ denote the set of users in the enterprise. To detect occurrences of malicious activity in the network, the authentication activity of each user $i \in V$ is monitored via hourly counts of network logons per source computer. Let $M$ denote the number of distinct source computers in the network. For each user $i \in V$, let 
\begin{align}
\label{eq:datacyber}
x_{i, t} = (x_{i, t, 1}, \ldots, x_{i, t, M}),
\end{align}
where $x_{i, t, \ell}$ denotes the number of network logons initiated by user $i$ from source computer $\ell$ during the $t$-th hour of the $58$ day data collection period. For each user, it is of interest to detect temporal changes in the distribution of network logons across source computers as possible evidence for malicious activity. Figure \ref{fig:motiv1} in Appendix \ref{sec:appendixcyber} %the supplementary material \citep{SuppCPGTS} 
displays the authentication data for two users. % 

\vspace{-3mm}
\subsection{Motivation for an informative graph-based changepoint prior} %Encoding graph-based prior information}
\label{sec:graphusers}

The authentication data (\ref{eq:datacyber}) exhibit much variability, and some observed changes can correspond to legitimate activity. Therefore, to limit the number of false alerts and yet not overlook weak signals from genuine attack footprints, it is key to incorporate prior knowledge in the change detection procedure.

When attackers penetrate a network, they rarely gain access to the target users directly; instead, they typically take control of a vulnerable user, for example via email phishing, and then they move laterally through the network, gaining access and compromising additional users, to achieve their objectives \citep{Sexton2015AttackCD}. Attackers are typically constrained in the way they can navigate the network, and it will often be possible for cyber-security experts to specify a graph $G = (V, E)$, where an edge $(i, i^\prime) \in E \subseteq V \times V$ indicates it is believed \textit{a priori} that attackers may 
switch credentials 
% move 
between user $i$ and user $i^\prime$ at any time during the data collection period. 
%  Attackers may \textit{a priori} move between user $i$ and user $i^\prime$ to traverse the network 
Therefore, it is of interest to encode in the changepoint prior that cyber-attacks are \textit{a priori} likely to result in quasi-synchronous changes in the authentication activity of multiple users that are connected in $G$. In this article, we consider the following specification of $G$ for demonstration purposes:  $(i, i^\prime) \in E$ if and only if both user $i$ and user $i^\prime$ successfully initiated a network logon from the same source computer on the same day. This choice follows from the following considerations. 
% To move from some user $i$ to some user $i^\prime$, 
%
In Windows operating systems, when a user logs on with their credentials  (username and password hash) to a device on the domain, these credentials are cached locally on the device. Credential caching prevents users from continuously having to re-authenticate (single sign-on), and enables them to log on to the device even if the device is disconnected from the network. Attackers will exploit credentials which are cached on devices to upgrade their privileges and move laterally through the network. How long credentials may be cached on devices depends on the enterprise's network settings. In the absence of precise knowledge about the enterprise's network settings, it is reasonable to assume that if both user $i$ and user $i^\prime$ have logged into a device on the same day then both those credentials may be cached on that device during the data collection period. As a result, if attackers had access to that device then they would have the ability to exploit cached credentials to switch credentials between user $i$ and user $i^\prime$.

\begin{figure}[t!]
\centering
\vspace{-2mm}
\scalebox{.75}{
\includestandalone[width=1.0\textwidth]{networkmotiv}%
}
    \caption{Cartoon representation of a cluster of synchronous changepoints (red crosses) on a graph of time series. Arrows represent time series, and shaded rectangles indicate which pairs of time series are likely to be impacted by simultaneous changepoints. % connections between time series). 
    }
        \label{fig:motivcyber}
    \vspace{-5mm}
\end{figure}

In Figure \ref{fig:motivcyber}, for the application of interest, each arrow corresponds to the authentication activity of a user on the network, and shaded rectangles indicate which pairs of users are connected in $G$ and therefore likely to be impacted by simultaneous changes during an attack. It is of interest to encode in the changepoint prior, by means of the graph $G$, % which pairs of users 
that pairs of users $(i, i^\prime) \in E$ 
are likely to be simultaneously affected by malicious behavioural changes, thereby inducing a changepoint model for the authentication data that borrows strength across connected users in $G$ to detect signals for clusters of synchronous changes, as sketched in Figure \ref{fig:motivcyber}.

 In contrast with recent intrusion detection methods \citep{ChenZhang, chen2019a, SilviaNewEdge, passino2021graph}, the focus of this article is not the temporal evolution of a graph representing a network, and both $V$ and $E$ are constant in time. The graph $G$ represents the best available static characterisation of the network that can be used to guide change detection in the authentication activity of users (\ref{eq:datacyber}), and it is assumed to be readily available prior to running network intrusion detection methods; note that, in practice, the edge set could be derived from historic data.  Section \ref{sec:discuss} discusses possible model extensions for settings where prior beliefs on which time series are likely to be impacted by simultaneous changepoints may be time-dependent.

\section{Changepoint analysis for multiple time series}
\label{sec:changepointanalysisoverview}

Let $G=(V, E)$ be a graph with node set $V = \{1, \ldots, N\}$ and edge set  $E \subseteq V \times V$. For each node $i \in V$ we observe a time series $\boldsymbol{x}_i = (x_{i,0}, \ldots, x_{i,T})$ which may be subject to changepoints, and the edge set $E \subseteq V \times V$ indicates which pairs of time series are \textit{a priori} likely to be impacted by quasi-simultaneous changepoints. 
Conditionally on changepoints, the data %$\boldsymbol{x} = (\boldsymbol{x}_i  )_{i \in V}$ 
are assumed to be independent of $G$ and follow a standard parameteric changepoint model, presented in this section. Some limitations of the usual prior for independent changepoints are discussed, paving the way for %the introduction of 
the proposed informative prior for graph-dependent changepoints. % across time series in $G$. % which induces a graphical changepoint model for the data  $\boldsymbol{x} = (\boldsymbol{x}_i  )_{i \in V}$.

\subsection{Model and notation}
\label{sec:generalmodel}

For each node $i \in V$, suppose there are $k_i \geqslant 0$ changepoints that partition the time series of observations for that node into $k_i +1$ segments. 
% the passage of time into $k_i +1$ segments. 
The ordered locations of the changepoints, denoted by $\boldsymbol{\tau}_i = (\tau_{i,1}, \ldots,  \tau_{i, k_i})$, belong to the set $\mathcal{T}_{k_i}$, where
\begin{align}
\mathcal{T}_{k} = \left\lbrace (\tau_{1}, \ldots,  \tau_{k} ) \in \mathbb{N}^{k}; \:  0 \equiv  \tau_{0} < \tau_{1} < \cdots < \tau_{k} < \tau_{k +1} \equiv T+1\right\rbrace.
\label{eq:deftau}
\end{align}
For each  node $i$, the data $x_{\tau_{i, j-1}}, \ldots, x_{\tau_{i, j}-1}$ in each segment $j$ are assumed to be drawn from a distribution from the same parametric family $L_i(\cdot | \theta_{i, j})$, with a segment specific parameter $\theta_{i, j}$ drawn independently from a prior density $\pi_{i}(\cdot)$.

The parameters of interest are the changepoint parameters $(\boldsymbol{k}, \boldsymbol{\tau})$, where $\boldsymbol{k} = (k_i)_{i \in V}$ and $\boldsymbol{\tau} = (\boldsymbol{ \tau}_{i})_{i \in V}$.  Motivated by computational considerations, as in \citet{exactcpt} it is assumed in this article that segment parameters may be marginalised so that the likelihood of the data $\boldsymbol{x}$ conditional on changepoints, 
\begin{align}
\label{eq:likelihoodfull}
\mathcal{L}(  \boldsymbol{x} | \boldsymbol{k}, \boldsymbol{\tau}  ) = \prod_{i\in V} \prod_{j=1}^{k_i+1} \mathcal{L}_{i}( \tau_{i,j-1}, \tau_{i,j}),
\end{align}
%$L(  \boldsymbol{x} | \boldsymbol{k}, \boldsymbol{\tau}  ) = \prod_{i\in V} \prod_{j=1}^{k_i+1} \mathcal{L}_{i}( \tau_{i,j-1}, \tau_{i,j})$,
where 
\begin{align}
\label{eq:likelihood}
\mathcal{L}_{i}( \tau_{i,j-1}, \tau_{i,j}) = \int L_i(  x_{i,  \tau_{i,j-1}},  \ldots,  x_{i, \tau_{i,j}-1}| \theta_{i, j} ) \pi_i(\theta_{i, j} ) d \theta_{i, j}
\end{align}
%for all $t_1 < t_2$,  
can be computed. Given a prior for the changepoint parameters, $\pi( \boldsymbol{k}, \boldsymbol{\tau} ) $, one can consequently compute the posterior density function for the changepoint parameters, up to a normalising constant.

Examples of changepoint models where segment parameters may be marginalised 
%Examples of changepoint models satisfying the above assumptions 
include models for independent and identically distributed data within segments \citep{exactcpt, Denison}, 
changing linear regressions \citep{Punskaya, HieraCPCarlin}, models for time-dependent data within segments, such as Markov models with time-varying transition matrices \citep{Bolton2018}, zero-mean and heteroscedastic processes with changing variance \citep{JohnsonCP2003}, and changepoint models with segment parameters subject to seasonal effects \citep{TurcottePhD}.   
%Moreover, some model extensions where segment parameters cannot be marginalised, or where segment parameters are not independent,  are  discussed in Section \ref{sec:inferenceMCMC}, indicating how the proposed sampling strategy could be adapted for these model extensions. 
Moreover, some model extensions where segment parameters cannot be marginalised, and where segment parameters may be shared across segments, are discussed in Appendix \ref{sec:possiblemodelextension}, % in the supplementary material \citep{SuppCPGTS}, 
indicating how the proposed sampling strategy could be adapted for these model extensions.

In particular, consider the class of changepoint models where, within each segment, the data are assumed to be independent and identically distributed such that 
\begin{equation}
\label{eq:classicconiid}
x_{i,t} \overset{ }{\sim} f_i( \, \cdot \, | \,  \theta_{i,j}), \quad   \tau_{i, j-1} \leqslant t < \tau_{i, j},
\end{equation}
for some parametric density $f_i( \, \cdot \, | \,  \theta_{i, j})$ dependent on some segment parameter $\theta_{i, j} \sim \pi_i(\cdot)$.  
The integrals in (\ref{eq:likelihood}) can be calculated analytically when $\pi_i$ is chosen to be conjugate to $f_i$; and for non-conjugate cases, (\ref{eq:likelihood}) may be calculated numerically for low-dimensional segment parameters.
For the cyber-security application discussed in Section \ref{sec:motivationalsection}, the changepoint model (\ref{eq:classicconiid}) is suitable for the count data 
 %(\ref{eq:datacyber}) 
 with, for all $i$, $f_i$ denoting the density of the multinomial distribution with unknown probability parameter vectors $\theta_{i, j}$ with an uninformative, conjugate prior $\text{Dirichlet}( \mathbf{1}^{M} )$, where $\mathbf{1}^{M}$ denotes the $M$-dimensional vector of ones. As a result, each changepoint $\tau_{i, j}$ corresponds to a temporal change in the distribution of counts of  logons initiated by the user $i \in V$ across $M$ host computers in the network.

\vspace{-2mm}

\subsection{Limitations of the standard prior for independent changepoints}

When changepoints are assumed to be independent across time series, the posterior distribution of changepoints can be estimated for each time series separately. In this setting, it is standard to assume \textit{a priori} that, for all time series, discrete time changepoints follow a Bernoulli process \citep{exactcpt} such that %, for all changepoint parameters $(\boldsymbol{k}, \boldsymbol{\tau})$,
\begin{align}
\pi(\boldsymbol{k}, \boldsymbol{\tau}  |  p ) = \prod_{i \in V} p^{k_i} (1-p)^{T-k_i}  \mathbb{1}_{ \mathcal{T}_i }( \boldsymbol{\tau}_i )
\label{eq:classicprior}
\end{align}
for some Bernoulli parameter $0< p <1$, which encodes prior belief on the expected number of changepoints.  %$\boldsymbol{p} = ( p_1, \ldots, p_{N})$ with $0< p_i <1$ for all $i$. 

For the cyber-security application where $G$ represents a network of users, the standard prior in (\ref{eq:classicprior}) cannot fully encode prior beliefs on changepoints. Appendix \ref{sec:appendixcyberstudy} % in the supplementary material \citep{SuppCPGTS} 
exposes limitations resulting from the assumption of changepoint independence across time series through a comparative study.
%
%Appendix expose limitations resulting from the assumption of changepoint independence across time series.
%
%To expose limitations resulting from the assumption of changepoint independence across time series. 
No choice of $p$ seems satisfactory: choosing a small value for $p$ will limit the number of false alerts due to noise in user-specific legitimate activity; yet it will also prevent the detection of weak signals for changes shared by different users which are linked in the network, that may be of great interest. It would be preferable to specify \textit{a priori} that changepoints are more likely to occur simultaneously across time series that are linked in $G$, in order to require strong evidence from the data for changes impacting a single user, or possibly weak signals for changes that impact multiple users  linked in the network.

\section{Graphical models for dependent changepoints across multiple time series}

This section proposes a novel graphical prior for dependent changepoints across multiple time series. %, which induce a graphical changepoint model for the data $\boldsymbol{x}$. 
Given the graph of time series $G = (V, E)$, where $V=\{1, \ldots, N\}$, changepoints are modelled by means of an undirected graphical model encoding that pairs of time series $(i, i^\prime) \in E$ are \textit{a priori} likely to be simultaneously affected by changepoints. The graphical model is further extended by relaxing the assumption that dependent changepoints across time series are synchronous; the extended model assumes dependent changepoints across time series correspond to nearby but not necessarily identical time points.

\label{sec:newprior}%

\subsection{Synchronous dependent changepoints across time series} 

\label{sec:markovrandomfield}

\setcounter{secnumdepth}{5}
\subsubsection{Model definition}

In Section \ref{sec:generalmodel}, changepoints were most simply defined in terms of  % changepoint parameters $(\boldsymbol{k}, \boldsymbol{\tau})$. 
their number and locations, $(\boldsymbol{k}, \boldsymbol{\tau})$. 
Subsequently, it will be useful to represent changepoints by means of a binary matrix. For changepoint parameters $(\boldsymbol{k}, \boldsymbol{\tau})$,  let $\boldsymbol{S} = (S_{i,t})$ 		% $\equiv \boldsymbol{S}(\boldsymbol{k}, \boldsymbol{\tau})$ 
be the corresponding binary matrix such that, for all $i \in V$ and $t=1, \ldots, T$,
\begin{align}
S_{i,t} = \left\{ \begin{array}{ll}
         1 & \mbox{if $\exists j \in \{1, \ldots, k_i \}  \text{ s.t. } t= \tau_{i, j} $}\\
        0 & \mbox{otherwise},
        \end{array} \right.
\label{eq:binaryrepresentation}
\end{align}
so that $(\boldsymbol{k}, \boldsymbol{\tau})$ and $\boldsymbol{S}$ are equivalent representations of the changepoints. Moreover, let $S_{i, 0} = S_{i, T+1} = 1$ for all $i$.

To encode the dependence structure of synchronous changepoints across time series in $G = (V, E)$, 
let $\boldsymbol{\lambda} = (\lambda_{i,i^{\prime} })$ be a symmetric matrix of non-negative edge weights for the graph satisfying $\lambda_{i,i^{\prime} } > 0 $ if and only if $(i, i^\prime) \in E$ for all $i,i^\prime \in V$. Then, conditional on $\boldsymbol{\lambda}$, changepoints are assumed to have a prior distribution described by the weighted, 
undirected graph $G$ 
such that, for all $(\boldsymbol{k}, \boldsymbol{\tau})$,
\begin{align}
\pi(\boldsymbol{k}, \boldsymbol{\tau} | p, \boldsymbol{\lambda} )   & = \frac{1}{Z(p, \boldsymbol{\lambda} )} \prod_{t=1}^T  \exp \left\lbrace  \bar{p}\sum_{i \in V}  S_{i,t} + \sum_{ i < i^{\prime}  } \lambda_{i, i ^\prime} S_{i,t} S_{i^\prime, t} \right\rbrace,
\label{eq:MRFS}
\end{align}
for some $0 < p < 1$, where $\bar{p}= \textnormal{logit}(p) =  \log\{ p/(1-p)\}$ and some  normalising constant $Z(p, \boldsymbol{\lambda} )$ that has no convenient closed form in general but will present no computational complications since the MCMC algorithm for changepoint parameters proposed in Section \ref{sec:inferenceMCMC} only requires computation of ratios of the prior density \eqref{eq:MRFS}.

If the edge set $E$ is the empty set, implying $\lambda_{i,i^\prime}=0$ for all $i$ and $i^\prime$, then the prior distribution in (\ref{eq:MRFS}) is equivalent to the standard prior for independent changepoints (\ref{eq:classicprior}); for all changepoint parameters $(\boldsymbol{k}, \boldsymbol{\tau})$ and for all $0<p<1$, % we have 
\begin{align}
\label{eq:standard0}
\pi(\boldsymbol{k}, \boldsymbol{\tau}| p,  \boldsymbol{0}) = \prod_{i \in V} p^{\sum_{t=1}^T S_{i, t}} (1-p)^{T-\sum_{t=1}^T S_{i, t}} ,
\end{align}
%\pi(\boldsymbol{k}, \boldsymbol{\tau} | p),
where
 $\boldsymbol{0}$ is the null matrix. 
 %$\boldsymbol{0}$ is the null matrix and $Z(p, \boldsymbol{0})= (1-p)^{-NT}$. 
 The memoryless property of the standard prior \citep{exactcpt} is maintained by the extended prior (\ref{eq:MRFS}), conditional on fixed value of $p$. The latter assumes independent and identical Markov random fields \citep{lauritzen1996} for the columns of $\boldsymbol{S}$. The memoryless property would be lost if (\ref{eq:MRFS}) were marginalised over a prior distribution for $p$. %we assumed a prior for $p$.

The graphical prior distribution (\ref{eq:MRFS}) takes into account both the number  of changepoints across time series and their relative positions; the parameter $p$ controls prior belief on the sparsity of changepoints, and the edge weight parameters $\boldsymbol{\lambda}$ control  
 the synchronisation of changepoints between time series.  
For all pairs $(i, i^\prime)$, the larger the edge weight $\lambda_{i, i^\prime} > 0$, the higher the probability for time series $i$ and $i^\prime$ to be simultaneously affected by changepoints.
%Hence, with the prior in (\ref{eq:MRFS}) it is possible to specify changepoints are likely to  occur simultaneously across clusters of time series according to $G$.
Hence, the prior in (\ref{eq:MRFS}) may specify changepoints are likely to  occur simultaneously across clusters of time series according to $G$.

To understand how to set the changepoint prior parameters $p$ and $\boldsymbol{\lambda}$ in practice, it is instructive to consider the conditional prior distribution of the components of the binary matrix $\boldsymbol{S}$. Under (\ref{eq:MRFS}), the conditional  distribution of $S_{i,t}$ given $\boldsymbol{S}_{-(i,t)} = \{ S_{i^\prime, t^\prime}: \, (i^\prime, t^\prime) \neq (i,t) \}$ is 
\begin{align}
\pi(S_{i,t} | \boldsymbol{S}_{-(i,t)}, p, \boldsymbol{\lambda}) \propto  \exp\left\lbrace  S_{i,t} \left( \bar{p} + \sum_{i^\prime: \, (i , i^\prime) \in E} \lambda_{i, i ^\prime} S_{i^\prime, t} \right)  \right\rbrace, \quad S_{i,t} \in \{0, 1\}.
\end{align}
Therefore, for all $i$ and $t$, the hyperparameter 
$p$ 
 %$p =  1/(1+\exp\{ - \bar{p}\})$
  corresponds to the prior probability that $t$ is a changepoint for the $i$th time series given that no changepoints occur at time $t$ for the graph neighbour time series of $i$; and, for all $i^\prime$ such that $(i, i^\prime) \in E$, the interaction parameter $\lambda_{i,i^\prime}$ governs how much the conditional prior probability increases if the neighbour time series $i^\prime$ is impacted by a changepoint at time $t$. 
Moreover, to perceive the influence of the changepoint  prior parameters on the posterior distribution of changepoints, it is helpful to consider the full conditional distribution of $S_{i,t}$ given $\boldsymbol{S}_{-(i,t)} = \{ S_{i^\prime, t^\prime}; \, (i^\prime, t^\prime) \neq (i,t) \}$, 
\begin{align}
\pi(S_{i,t} | \boldsymbol{S}_{-(i,t)}, p, \boldsymbol{\lambda}, \boldsymbol{x}) &\propto \left( \frac{\mathcal{L}_i(\tau_t^{\prime}, t)\mathcal{L}_i(t, \tau_t^{\prime \prime} ) }{\mathcal{L}_i(\tau_t^{\prime}, \tau_t^{\prime \prime})}  \right)^{S_{i,t}}  \pi(S_{i,t} | \boldsymbol{S}_{-(i,t)}, p, \boldsymbol{\lambda})   \nonumber \\
 &\propto \exp\left\lbrace  S_{i,t} \left( \log \left\lbrace \frac{\mathcal{L}_i(\tau_t^{\prime}, t)\mathcal{L}_i(t, \tau_t^{\prime \prime} ) }{\mathcal{L}_i(\tau_t^{\prime}, \tau_t^{\prime \prime})}  \right\rbrace +  \bar{p} + \sum_{i^\prime: \, (i , i^\prime) \in E } \lambda_{i, i ^\prime} S_{i^\prime, t} \right)  \right\rbrace,
\label{eq:fullconditionalS}
\end{align}
where $\mathcal{L}_i$ is defined in (\ref{eq:likelihood}), $\tau_t^{\prime} = \max\{ t^\prime:  \, t^\prime < t,   S_{i, t^\prime} = 1  \}$ and $\tau_t^{\prime \prime} =\min\{ t^\prime:  \, t^\prime > t, S_{i, t^\prime} = 1 \}$. 
In essence, $p$ determines the level of evidence required from the data to suggest a changepoint, 
%can be considered strong, 
and the edge weight parameters control, relative to $p$, how weak signals for synchronous changepoints can be combined across time series.

\subsubsection{A special case: identical edge weight parameters}
\label{sec:aspecialcase}
In practice, it will often be natural to assume that, for all $(i, i^\prime) \in E$, $\lambda_{i ,i^\prime} = \lambda$ for some fixed value $\lambda >0$. For all $i$ and $t$, 
let 
\begin{align}
n_{i, t} =  \sum_{i^\prime:  (i, i^\prime) \in E } S_{i^\prime, t} 
\label{eq:numneighb}
\end{align}
be the number of neighbour time series of $i$ that are affected by a changepoint at time $t$. Then, under (\ref{eq:MRFS}), the conditional prior distribution of $S_{i,t}$ given $\boldsymbol{S}_{-(i,t)} = \{ S_{i^\prime, t^\prime}: \, (i^\prime, t^\prime) \neq (i,t) \}$ is 
\begin{align}
\label{eq:fullcondspecial}
\pi(S_{i,t} | \boldsymbol{S}_{-(i,t)}, p, \lambda) = \frac{ \exp\left\lbrace  S_{i,t} \left( \bar{p} + \lambda n_{i, t} \right)  \right\rbrace }{\exp\left\lbrace  \bar{p} + \lambda n_{i, t}   \right\rbrace + 1}, \quad S_{i,t} \in \{0, 1\}.
\end{align}
Moreover, $\lambda$ will typically be chosen relative to $\bar{p}$ and the degree distribution of the nodes in $G$. 
For example, it can be convenient to assume $\lambda = \lambda_s  |\bar{p}|/n$, where $n$ denotes the maximum degree of the nodes in $G$, for some $\lambda_s > 0$.
% For example, it can be convenient to assume $\lambda \propto  |\bar{p}|/n$, where $n$ denotes the maximum degree of the nodes in $G$. 

\subsection{Examples of graphical dependence structures for changepoints}
\label{sec:dependencestructureseg}

The prior distribution (\ref{eq:MRFS}) is suitable for a wide variety of settings. This section provides graph motifs that can be regarded as building blocks to encode the dependence structure of changepoints
 across multiple time series. 
For these examples, we 
assume identical non-zero edge weights as considered 
% consider the parametrisation introduced 
in Section \ref{sec:aspecialcase} and provide some insight on how to choose the changepoint prior parameters $p$ and $\lambda$. 
%, hoping it gives intuition on how to set hyperparameters for more realistic graphs. % such as computer networks.  
These exemplar dependence structures for changepoints are explored via a simulation study in %Section \ref{appendix:simustudy}.
Appendix \ref{sec:appendixsimustudy}. % in the supplementary material \citep{SuppCPGTS}. 
%%%

\subsubsection{Lattices}
\label{sec:latticesegs}
It might be natural to choose the edge set $E$ to induce an $N_1 \times N_2 $ lattice graph when the number of time series is $N= N_1N_2$ for some $N_1, N_2 > 0 $. For all $i$, let $0 \leqslant i_1 \leqslant N_1 -1$ and $0 \leqslant i_2 \leqslant N_2-1$ be the unique natural numbers such that $i =  i_2 N_1 + i_1 + 1$. % with $0 \leqslant i_2 \leqslant N_2$. 
Then the $N_1 \times N_2 $ lattice graph is such that $(i, i^\prime) \in E$ if and only if $|i_1-i_1^\prime| + |i_2-i_2^\prime| = 1$. 
%Letting $i =  i_1 N_1 + i_2 + 1$ with $0 \leqslant i_2 \leqslant N_2$ for all $i$, then $(i, i^\prime) \in E$ if and only if $|i_1-i_1^\prime| + |i_2-i_2^\prime| = 1$. 
%
For example, suppose the data $\boldsymbol{x}$ are recorded for the analysis of some spatio-temporal phenomenon such that $x_{i,t}$ denotes the observation at time $t$ and at the coordinate $i$ of some $N_1 \times N_2 $ grid over a map of the region of interest, and it is of interest to detect the times and the coordinates at which the distribution of the data changes.

Figure \ref{fig:graphillustrateNEW} illustrates the dependence structure for changepoints induced by the graphical changepoint prior (\ref{eq:MRFS}) given a lattice graph $G$ on $20$ time series of length $T$. %for $20$ time series that lie in a $5 \times 4$ lattice graph $G$. 
The larger the edge weight $\lambda > 0$, the higher the probability for pairs of time series connected on the lattice graph $G$ to be simultaneously impacted by changepoints. As a result, in Figure \ref{fig:graphillustrateNEW}, changepoints at time $t^\prime$, which are connected by edges, are \textit{a priori} more likely than isolated changepoints at time $t$. The prior (\ref{eq:MRFS}) can therefore specify that changepoints are likely to occur as clusters of simultaneous changepoints on the lattice.  The conditional probability (\ref{eq:fullcondspecial}) specifies that $p$ is the prior probability that a changepoint occurs in isolation on the lattice, and is constrained such that $n_{i, t} \in \{0, \ldots, 4\}$.

\begin{figure}[t!]
\centering
%\vspace{-5mm}
\hspace{-5mm}
\scalebox{.9}{
\includestandalone[width=1.0\textwidth]{eggraphSmatrix}%
}
    \caption{Cartoon representation of a changepoint matrix $\boldsymbol{S} = (S_{i, t})$, defined in (\ref{eq:binaryrepresentation}), 
    for $20$ time series of length $T$ which lie on a $5 \times 4$ lattice graph $G$. Edges indicate dependence between components of $\boldsymbol{S}$ according to the prior (\ref{eq:MRFS}) given the lattice graph $G$.  Blue squares indicate $S_{i, t}=1$ and white circles indicate $S_{i, t}=0$ for all $i$ and $t$. % Edges indicate dependence between the components of $S$.
    }
        \label{fig:graphillustrateNEW}
    \vspace{-2mm}
\end{figure}

\vspace{-1mm}
\subsubsection{$r$-chains}
\label{sec:rchainsegs}
Another dependence structure of interest arises when there is a natural ordering of the time series, which is encoded by the time series indices $1 < \cdots < N$, and changepoints are \textit{a priori} likely to occur as chains of simultaneous changepoints across consecutive time series. For instance, suppose the data consist of multiple time series that are recorded to monitor various aspects of a system; 
and, it is of interest to detect some event which evolves through multiple phases, such that each phase is likely to manifest through the pertubation of one aspect of the system.
In such a setting, it is appropriate to consider the following graph for the time series indices, which we call an  $r$-chain graph:  let $(i, i^\prime) \in E$ if and only if $1 \leqslant |i - i^\prime| \leqslant r$, for some $r >0$ chosen to allow gaps of length $r-1$ within  chains of changepoints. 
%Panel (b) of Figure \ref{fig:graphsimul} gives a cartoon representation of a $2$-chain graph, which will be considered in a simulation study. 
For $r$-chain graphs, $n_{i, t} \in \{0, \ldots, 2r\}$. 
%For the prior given in (\ref{eq:MRFS}), $\bar{p}$ and $\lambda$ are chosen such that, for all $i$ and $t$, the prior conditional probability that $t$ is a changepoint for the $i$th time series is $1/(1 + \exp\{ -\bar{p} - n_{i, t} \lambda  \} )$ where $n_{i, t} \in \{0, \ldots, 2r\}$.

\vspace{-1mm}

\subsubsection{Complete graphs}
\label{sec:completegraph}
Suppose a complete graph for the time series indices, that is $(i, i^\prime) \in E$ for all $i \neq i^\prime$, so that, according to Section \ref{sec:aspecialcase},  $\lambda_{i, i^\prime} = \lambda > 0$ for all $i \neq i^\prime$.  %  the graph structure is superflous and 
In such a setting, the prior given in (\ref{eq:MRFS}) assumes changepoint locations are exchangeable across time series, like the MVCAPA model proposed in \citet{fisch}, %\citet{Jeng2012} and \citet{samworth}, 
and therefore solely takes into account the number of time series impacted by a changepoint at time $t$, for all $t$.

\subsubsection{Unknown graph}

This article assumes that the graph $G$ is known \textit{a priori} and contains useful information concerning the dependence structure of changepoints across time series. %to encode in the changepoint prior. 
Future work could reverse this idea and consider applications where  
estimating $G$ is one of the inferential objectives. 
%it might desirable to encode some uncertainty about the graph $G$. 
It might often be computationally unrealistic to specify an unconstrained prior for $G$ admitting that $\lambda_{i, j} \geqslant 0$ for all $(i, i^\prime) \in V \times V$. However, in some settings it might be appropriate to consider a class of possible graphs $G$, such as those considered in the previous two subsections; for example, it may be assumed \textit{a priori} that $G$ is an $r$-chain with $r \geqslant 0$ unknown.
%so that it is desirable to encode some uncertainty about the graph $G$. 

\subsection{Extension to asynchronous changepoint dependence}

\label{sec:asynccp}

The model in Section \ref{sec:markovrandomfield}  assumes changepoints are likely to simultaneously affect clusters of time series according to the dependence graph $G$. 
In this section, we relax this model to allow dependence between changepoints in different time series at nearby points in time. 
The extended model relies on representing changepoints as 
lagged realisations of simultaneous but unobserved 
 latent changepoints. The latent changepoints are distributed according to the model introduced in Section \ref{sec:markovrandomfield} and, conditional on these latent changepoints, time series-specific lags are assumed to be uniformly distributed over some small time window.

Let $(\boldsymbol{k}, \boldsymbol{\tau})$ be  changepoint parameters for multiple time series as defined in Section \ref{sec:generalmodel}, where it is assumed that the $i$th time series is subject to $k_i$ changepoints whose positions are denoted $\boldsymbol{\tau}_i =  (\tau_{i, 1}, \ldots, \tau_{i, k_i} ) \in \mathcal{T}_{k_i}$ as defined in (\ref{eq:deftau}). 
The asynchronous model further assumes that, for all time series $i=1, \ldots, N$, there exist 
 latent changepoint positions  $\boldsymbol{ \tilde{\tau} } = ( \boldsymbol{ \tilde{\tau} }_1, \ldots, \boldsymbol{ \tilde{\tau} }_N  )$, 
$\boldsymbol{ \tilde{\tau} }_i = (\tilde{\tau}_{i, 1}, \ldots, \tilde{\tau}_{i, k_i} ) \in \mathcal{T}_{k_i}$,  and  lags $\boldsymbol{d} = ( \boldsymbol{d}_1, \ldots, \boldsymbol{d}_N )$, 
$\boldsymbol{d}_i = ( d_{i, 1}, \ldots, d_{i, k_i} ) \in \{0, \ldots, w_i \}^{k_i} $, for  $\boldsymbol{w} = ( w_1, \ldots, w_N )$, where $w_i \geqslant 0$ is an upper bound for the lags, % and 
such that, for all $j=1, \ldots, k_i$, the $j$th changepoint for time series $i$ is
\begin{align}
\tau_{i,j} = \tilde{\tau}_{i,j}+ d_{i, j}.
\label{eq:latentc}
\end{align}
Let $\tilde{\tau}_{i,0} =\tau_{i,0} = 0$ and $\tilde{\tau}_{i,k_i+1} = \tau_{i,k_i+1} = T+1$. 
For all $(k_i, \boldsymbol{\tau}_i)$ and $w_i \geqslant 0 $, if $\boldsymbol{ \tilde{\tau} }_i = \boldsymbol{ \tau}_i$ and $\boldsymbol{d}_i $ is the zero vector then (\ref{eq:latentc}) holds, and therefore the existence of a corresponding pair $(\boldsymbol{ \tilde{\tau} }_i, \boldsymbol{d}_i )$ with $\boldsymbol{ \tau }_i \in \mathcal{T}_{k_i}$ is guaranteed. 
If $w_i=0$ then the latent changepoints and the changepoints must be identical; but, in general, given  changepoints $(k_i, \boldsymbol{ \tau}_i)$ and $w_i > 0$, there are multiple distinct pairs of latent changepoints $(k_i,  \boldsymbol{ \tilde{\tau} }_i)$ and lags $\boldsymbol{d}_i$ satisfying (\ref{eq:latentc}) and $\boldsymbol{ \tilde{\tau} }_i \in \mathcal{T}_{k_i}$.

For some applications the upper bounds for the lags, $\boldsymbol{w}$, may be fixed.  In particular, for some reference time series $i \in V$, it can be set that $w_i=0$, implying that $\tau_{i, j} = \tilde{\tau}_{i, j}$ for all $j$, so that changepoints for time series $i^\prime $ with $w_{i^\prime} \geqslant 0$ are lagged relative to changepoints for time series $i$. However, in general, upper bounds for the lags will not be known. For example, in the motivational application in cyber-security, no user $i \in V$ can be assumed to be the first user to be affected by an attack, making it awkward to pick a reference time series $i$, and the exact duration of attacks is not known \textit{a priori}. It will be assumed that, independently for all time series $i$, $w_i \sim \text{Geometric}(\eta)$ for some value $0 < \eta < 1$ chosen to reflect the expected duration of an attack.

Suppose the latent changepoints $(\boldsymbol{k}, \boldsymbol{\tilde{\tau} } )$ 
are distributed according to the prior distribution (\ref{eq:MRFS}) given some $0<p <1$ and graph edge weight parameters $\boldsymbol{\lambda}$. Then, independently for all time series $i$, conditional on $w_i$ and $(k_i,  \boldsymbol{\tilde{\tau} }_i)$, the lags $\boldsymbol{d}_i = (d_{i,1}, \ldots, d_{i,k_i})$ are assumed to be uniformly distributed on the set
\begin{align}
\mathcal{D}(w_i, k_i,  \boldsymbol{\tilde{\tau}}_i) = \left\lbrace ( d_{i, 1}, \ldots, d_{i, k_i} ) \in \{0, \ldots, w_i \}^{k_i} : \quad (\tilde{\tau}_{i,1} + d_{i, 1}, \ldots, \tilde{\tau}_{i,k_i} + d_{i, k_i}) \in \mathcal{T}_{k_i} \right\rbrace,
\label{eq:dsetlags}
\end{align} 
such that, for all $\boldsymbol{d}= (\boldsymbol{d}_1, \ldots, \boldsymbol{d}_N )$,
\begin{align}
\pi( \boldsymbol{d}  | \boldsymbol{k}, \boldsymbol{\tilde{\tau}}, \boldsymbol{w}  ) = \prod_{i=1}^{N}  \frac{ \mathbb{1}_{   \mathcal{D}(w_i, k_i,  \boldsymbol{\tilde{\tau}}_i) } (  \boldsymbol{d}_i )  }{  \text{card}(\mathcal{D}(w_i, k_i,  \boldsymbol{\tilde{\tau}}_i) )}  .
\label{eq:priorlags}
\end{align}
Proposition \ref{prop:cardinalityd} gives a recursion to derive the cardinality of (\ref{eq:dsetlags}).

\begin{propr}{\emph{(Cardinality of $\mathcal{D}$).}}
\label{prop:cardinalityd}
Let $w \geqslant 0$, $k \geqslant 0$, $\boldsymbol{\tau} = (\tau_1, \ldots, \tau_k) \in \mathcal{T}_k$ with $\tau_0 = 1$ and $\tau_{k+1} = T+1$, and let $\mathcal{D}(w, k, \boldsymbol{\tau} ) $ be the set defined in (\ref{eq:dsetlags}).  
\begin{itemize}
\item[(i)] For all $j\geqslant 1$ and $ l \geqslant 0$, let $\rho(j,l) = \min\{w + 1,T+1 - \tau_j \} - (\tau_{j+l} - \tau_{j})$ and 
\begin{align}
Q(j, l) = \frac{ (  \rho(j,l) + l  )! }{(\rho(j,l)-1)! (l + 1)! } \mathbb{1}_{ \{ 0, \ldots, w \}  }(\tau_{j+l} - \tau_{j}). 
\label{eq:defQ}
\end{align}
%Q(j, l) = \frac{\Gamma(  \rho(j,l) + l + 1  )}{\Gamma(\rho(j,l)) \Gamma(l + 2) } \mathbb{1}_{ \{ 0, \ldots, w \}  }(\tau_{j+l} - \tau_{j}). 
Additionally, let $Z(0) = 1$, $Z(1) = Q(1, 0)$ and, recursively for all $k>1$, 
\begin{align}
Z(k) = \sum_{j = 1}^{k}  (-1)^{k-j}  Z(j-1)Q( j, k -j ).
\end{align}
Then 
\begin{align}
\mathrm{card}(\mathcal{D}(w, k, \boldsymbol{\tau} ) ) = Z(k).
\label{eq:Dcard}
\end{align}
In particular, if $k=0$ then $\boldsymbol{\tau}$ is the empty sequence and $\mathcal{D}(w, k, \boldsymbol{\tau} )$ contains a unique element, namely the empty sequence.
\item[(ii)]  $\mathrm{card}(\mathcal{D}(w, k, \boldsymbol{\tau} ) ) \leqslant (w+1)^k$ and the equality holds if and only if $\tau_{j+1} -\tau_j > w$ for all $j$. 
\end{itemize}
\begin{proof}
See Appendix \ref{sec:appendixproof}. % in the supplementary material \citep{SuppCPGTS}. 
\end{proof}
\end{propr}

Consequently, the joint prior density for $(\boldsymbol{k}, \boldsymbol{\tilde{\tau}}, \boldsymbol{d})$ is 
\begin{align}
\label{eq:priorjointlatentlags}
\pi( \boldsymbol{k}, \boldsymbol{\tilde{\tau} }, \boldsymbol{d} | p, \boldsymbol{\lambda}, \boldsymbol{w}) = \frac{  \pi( \boldsymbol{k}, \boldsymbol{\tilde{\tau} } | p, \boldsymbol{\lambda} )  }{  \prod_{i=1}^N \mathrm{card}( \mathcal{D}(w_i, k_i, \boldsymbol{\tau}_i ) )  }
\end{align}
and the induced changepoint prior distribution for $(\boldsymbol{k}, \boldsymbol{\tau} )$ is
\begin{align}
\pi(\boldsymbol{k}, \boldsymbol{\tau } |  p, \boldsymbol{\lambda}, \boldsymbol{w} ) = \sum_{ ( \boldsymbol{\tilde{\tau}}, \boldsymbol{d}) \in \Upsilon(\boldsymbol{k}, \boldsymbol{\tau}, \boldsymbol{w}  )} \pi( \boldsymbol{k}, \boldsymbol{\tilde{\tau} }, \boldsymbol{d} | p, \boldsymbol{\lambda}, \boldsymbol{w}) ,
\label{eq:priorextended}
\end{align}
where $\Upsilon(\boldsymbol{k}, \boldsymbol{\tau}, \boldsymbol{w}  )$ denotes the set of pairs of latent changepoints and lags, $(\boldsymbol{k}, \boldsymbol{\tilde{\tau}}, \boldsymbol{d})$, that identify the changepoints $(\boldsymbol{k},\boldsymbol{\tau})$ according to (\ref{eq:latentc}). 
%If no lags are allowed, that is $w_i=0$ for all $i$, then $\Upsilon(\boldsymbol{k}, \boldsymbol{\tau}, \boldsymbol{w}  ) = \{ ( \boldsymbol{k}, \boldsymbol{\tau}, \boldsymbol{0} )\}$ and the prior in (\ref{eq:priorextended}) is identical to the prior discussed in Section \ref{sec:markovrandomfield}. 

\section{Markov chain Monte Carlo inference}

\label{sec:inferenceMCMC}

%
%However, joint sampling of changepoints across time series is required 
Joint sampling of changepoints across time series is required 
when the assumption of independence for changepoints is relaxed. In this section, we propose a reversible jump MCMC algorithm \citep{Green1995} to sample changepoints in multiple time series, $(\boldsymbol{k}, \boldsymbol{\tau})$. 
The changepoint parameter space is augmented with auxiliary variables \citep{besagauxi, higdon} that induce clusters of time series indices according to $G$. Then, the reversible jump MCMC algorithm of  \citet{Denison} is extended to sample from the augmented parameter space, thereby providing a means to efficiently explore the changepoint parameter space. At each iteration of the algorithm, a new cluster of changepoints may be proposed or an existing cluster of changepoints may be deleted or shifted. The validity of the proposed MCMC algorithm follows immediately from the reversibility of the proposed moves. 
\textcolor{corrections}{Appendix \ref{sec:appendixMCMCalgo} % in the supplementary material \citep{SuppCPGTS} 
gives some indications on the time complexity of the algorithm and discusses possible extensions for settings where segment parameters cannot be marginalised.}   
 For notational simplicity it is assumed there are no missing data, but  even with data which are not independent and identically distributed within segments, any missing observations  would present no methodological complication, since missing data can be sampled from their predictive distribution within the proposed MCMC scheme \citep{gelmanbda04}.

\subsection{Sampler for synchronous dependent changepoints}
\label{sec:binarysampler}
We begin by proposing an MCMC algorithm to sample from the posterior distribution of changepoints when changepoint parameters are \textit{a priori} distributed according to the prior introduced in Section \ref{sec:markovrandomfield}, $\pi( \boldsymbol{k}, \boldsymbol{\tau}| p, \boldsymbol{\lambda} )$, given $0<p<1$ and some interaction parameters $\boldsymbol{\lambda}$.

To sample changepoints for multiple time series, consider the following adaptation of the standard MCMC algorithm to sample changepoints for a unique time series \citep{Denison}, which will be called the ``single site updating'' MCMC algorithm thereafter. At each iteration of the algorithm, with $(\boldsymbol{k}, \boldsymbol{\tau})$ denoting the latest particle of the sample chain, propose one of the following two moves: for a uniformly chosen index $(i, t)$, 
propose $S_{i,t}$ to be updated to $1-S_{i,t}$,
thereby allowing birth or death of a changepoint; alternatively, the position of a randomly chosen changepoint, $\tau_{i, j}$, is sampled uniformly from $\{\tau_{i, j-1} +1, \ldots,  \tau_{i, j+1} -1 \}$. 

%As a result of the graph interaction parameters $\boldsymbol{\lambda}$, 
With the graphical prior distribution (\ref{eq:MRFS}), 
synchronous changepoints can be correlated across time series, and the single site updating MCMC algorithm 
can become impractical, as illustrated in Appendix \ref{sec:simuauxi} % \textcolor{corrections}{in the supplementary material \citep{SuppCPGTS}} 
through a simulation study. Instead, it will be necessary to propose moves that allow birth, death or shift of clusters of synchronous changepoints according to the graph induced by $\boldsymbol{\lambda}$.

\subsubsection{Augmenting the parameter space with auxiliary variables}

\label{sec:augmentparam}

To provide a means of moving efficiently through the state space of the changepoint parameters, the parameter space is augmented with binary 
auxiliary variables $\boldsymbol{u} = (\boldsymbol{u}_1, \ldots, \boldsymbol{u}_T)$ such that, for all $t$, $\boldsymbol{u}_t$ is  
an $N \times N$ symmetric binary graph adjacency matrix with $(i, i^\prime)$ element $u_{t}(i, i^\prime)$. 
For all $t$, the prior density of $\boldsymbol{u}_t$ is assumed to take the conditionally independent form
\begin{align}
\pi( \boldsymbol{u}_t | \boldsymbol{\lambda}, \delta, \boldsymbol{k}, \boldsymbol{\tau}) =  \prod_{i < i^\prime}  q_t(i,i^\prime)^{u_{t}( i, i^\prime)} \{ 1- q_t(i,i^\prime) \}^{1-u_{t}( i, i^\prime)},
\label{eq:condiaux}
\end{align}
where
\begin{align}
q_t(i,i^\prime) = 1-\exp\{ -\delta \lambda_{i, i^\prime} (1-| S_{i,t} - S_{i^\prime, t} |)  \}
\label{eq:probq}
\end{align}
is the conditional probability that $u_{t}( i, i^\prime) =1 $, given a partial decoupling parameter $ \delta \geqslant  0$  
\citep{higdon} whose role will be discussed in Section \ref{sec:mcmcalgorithmintro}.
After observing data $\boldsymbol{x}$ distributed according to (\ref{eq:classicconiid}), the joint posterior density of the augmented parameters $(\boldsymbol{k}, \boldsymbol{\tau}, \boldsymbol{u})$ is 
\begin{align}
\label{eq:jointaugmented}
\pi( \boldsymbol{k}, \boldsymbol{\tau}, \boldsymbol{u} | p, \boldsymbol{\lambda}, \boldsymbol{x}, \delta) =  \pi( \boldsymbol{u} | \boldsymbol{\lambda}, \delta, \boldsymbol{k}, \boldsymbol{\tau}) \pi( \boldsymbol{k}, \boldsymbol{\tau} | p, \boldsymbol{\lambda}, \boldsymbol{x}),
\end{align}
where 
\begin{align}
\pi( \boldsymbol{u} | \boldsymbol{\lambda}, \delta, \boldsymbol{k}, \boldsymbol{\tau}) = \prod_{t=1}^T \pi( \boldsymbol{u}_t | \boldsymbol{\lambda}, \delta, \boldsymbol{k}, \boldsymbol{\tau}).
\end{align}

For all $t$, consider the graph $H_t = (V, E_t)$ with vertex set $V = \{1, \ldots, N\}$ and edge set $E_t$ such that $(i, i^\prime) \in E_t$ if and only if $u_{t}( i, i^\prime)= 1$ for all $i, i^\prime \in V$. According to (\ref{eq:probq}), if $u_{t}( i, i^\prime)= 1$ then $q_t(i,i^\prime) > 0$ and, consequently, $S_{i, t} = S_{i^\prime, t}$. As a result, with $\mathcal{C}_t$ denoting the set of connected components of $H_t$, for all clusters of time series $\gamma \in \mathcal{C}_t$, $S_{i, t} = S_{i^\prime,t}$ for all $i, i^\prime \in \gamma$. In other words, the auxiliary variables 
$\boldsymbol{u}_t$  
%$\boldsymbol{u}_t = (u_{t}(i, i^\prime) )$ 
induce a partition of the time series, $\mathcal{C}_t$, such that, for each cluster $\gamma \in \mathcal{C}_t$, either all time series or no time series in $\gamma$ are affected by a changepoint at time $t$. Moreover, according to (\ref{eq:probq}), if $S_{i,t} = S_{i^\prime, t}$ then the conditional probability that $u_{t}(i, i^\prime)=1$ %, which is equal to $1- \exp\{ -\delta \lambda_{i, i^\prime} \}$, 
increases with $\lambda_{i, i^\prime} >0$, so that clusters induced by $\boldsymbol{u}_t$ will tend to be clusters on the graph induced by the edge weight parameters.

\subsubsection{MCMC algorithm}
\label{sec:mcmcalgorithmintro}
To generate realisations from the posterior distribution of the changepoints, we consider a ``cluster updating'' MCMC algorithm  that samples from the extended joint posterior density (\ref{eq:jointaugmented}). By inducing clusters of time series determined by the edge weight parameters for each time point $t$, the auxiliary variables $\boldsymbol{u}$ provide a means to efficiently explore the state space of $(\boldsymbol{k}, \boldsymbol{\tau})$. 

The parameter $\delta$ is a tuning parameter for the cluster updating MCMC algorithm. The size of clusters will tend to increase with $\delta$; in particular, if $\delta = 0$ then, for all $t$, each cluster corresponds to a unique time series index, even if the edge weights of the dependence graph are large, so that the ``cluster updating'' MCMC algorithm reduces to the ``single site updating'' algorithm. Typically $\delta$ is fixed \citep{higdon} to control the probabilities in (\ref{eq:condiaux}) and therefore the expected size of clusters. However, we propose to treat $\delta \geqslant 0$ as an unknown parameter with prior distribution $\pi(\delta)$, so that expected sizes of cluster may vary in the sample; specifically, we assume that $\delta = 0$ with probability $0 \leqslant \delta_0 \leqslant 1$ and otherwise $\delta$ is drawn from $\text{Beta}(\delta_1, \delta_2)$ for $\delta_1, \delta_2 > 0$.

For the cluster updating MCMC algorithm, at each iteration of the algorithm, with 
$(\boldsymbol{k}, \boldsymbol{\tau}, \boldsymbol{u}, \delta)$ 
denoting the latest particle of the sample chain, one of the following moves is proposed.

\subsubsection*{Birth/death move} 
Conditional on the auxiliary variables, the birth/death move proposes the birth or death of a cluster of synchronous changepoints. 
Sample $t^\prime$ uniformly from $\{ 1, \ldots, T\}$.  
A cluster $\gamma$ of time series indices is randomly chosen from $\mathcal{C}_{t^\prime}$, the set of clusters of time series induced by the auxiliary variables  
$\boldsymbol{u}_{t^\prime}$. 
%$\boldsymbol{u}_{t^\prime} = (u_{t^\prime}(i, i^\prime) )$. 
Then, leaving the auxiliary variables unchanged, propose changepoint parameters $(\boldsymbol{k}^\prime, \boldsymbol{\tau}^\prime)$ such that, for all $i = 1, \ldots, N$  and $t=1, \ldots, T$, 
\begin{align}
S_{i, t}^{\prime}  = \left\{ \begin{array}{ll}
         1- S_{i, t} & \mbox{if $i \in \gamma$ and $t =t^\prime$ }\\
        S_{i, t} & \mbox{otherwise},
        \end{array} \right.
\end{align}
where $\boldsymbol{S}$ and $\boldsymbol{S}^\prime$ are binary matrix representations of $(\boldsymbol{k}, \boldsymbol{\tau})$ and $(\boldsymbol{k}^\prime, \boldsymbol{\tau}^\prime)$ according to (\ref{eq:binaryrepresentation}), respectively.

\subsubsection*{Shift move}
The shift move proposes to shift the position of a cluster of synchronous changepoints. First, a time unit $t$ is uniformly chosen from $\{ t: \sum_{i=1}^N S_{i, t} > 0 \}$. 
Let $\mathcal{C}_t^{*} \subseteq \mathcal{C}_t$ denote the set of clusters of time series indices $\gamma$ induced by $\boldsymbol{u}_t$ such that, for all $i\in \gamma$,  $S_{i, t} = 1$. %$t \in \boldsymbol{\tau}_i$. 
A cluster $\gamma$ is uniformly chosen from $\mathcal{C}_t^{*}$.
For all $i \in \gamma$, let $j_i$ be the index of the changepoint with position $t$ for the $i$th time series, that is $\tau_{i, j_i}=t$. Then, sample uniformly $t^\prime$ from $\bigcap_{i \in \gamma} \{ \tau_{i, j_i -1} +1, \ldots, \tau_{i, j_i +1} -1 \}$ and propose changepoint parameters $(\boldsymbol{k}, \boldsymbol{\tau}^\prime )$ that are identical to  $(\boldsymbol{k}, \boldsymbol{\tau} )$ but with $\tau_{i, j_i} ^\prime =t^\prime$, for all $i \in \gamma$.

In parallel, it is required to propose auxiliary variables $\boldsymbol{u}^\prime$ which are adapted to $(\boldsymbol{k}, \boldsymbol{\tau}^\prime )$. The updated auxiliary variables differ from $\boldsymbol{u}$ as follows. For all $i, i^\prime \in \gamma$, $u_{t^\prime}^\prime(i,i^\prime) = u_{t}(i,i^\prime)$ and $u_{t}^\prime(i,i^\prime) = u_{t^\prime}(i,i^\prime)$; for all $i \in \gamma$ and $i^\prime \notin \gamma$ such that $t^\prime \notin \boldsymbol{\tau}_{i^\prime}$, $u^\prime_{t^\prime}(i,i^\prime) =0$; and, ensuring reversibility of the move, for all $i \in \gamma$ and $i^\prime \notin \gamma$ such that $t \notin \boldsymbol{\tau}_{i^\prime}$,  $u_{t}^\prime(i,i^\prime)$ is sampled conditionally on $(\boldsymbol{k}, \boldsymbol{\tau}^\prime)$ according to the $\text{Bernoulli} \left( 1- \exp \{ -\delta \lambda_{i, i^\prime} (1-| S_{i, t} - S_{i^\prime, t} |)  \} \right)$ target distribution implied by (\ref{eq:condiaux}).

\subsubsection*{Update of auxiliary variables}
Changepoints are left unchanged, $\delta$ is sampled from its prior distribution, and auxiliary variables are sampled from the full conditional distribution given in (\ref{eq:condiaux}), thereby proposing an updated clustering of time series indices for all $t$.

\subsection{Sampler for asynchronous dependent changepoints}
\label{sec:mcmcwithlags}
According to the changepoint model (\ref{eq:priorextended}) introduced in Section \ref{sec:asynccp}, changepoints do not need to occur at the same time to be related. Consequently, to explore the changepoint parameter space it will be required to propose the birth, death or shift of clusters of asynchronous changepoints. This section extends the MCMC algorithm from Section \ref{sec:binarysampler} to sample from the posterior distribution of changepoints when changepoint parameters are \textit{a priori} distributed according to $\pi(\boldsymbol{k}, \boldsymbol{\tau } |  p, \boldsymbol{\lambda}, \boldsymbol{w} )$ from (\ref{eq:priorextended}).

Recall that under the asynchronous model, changepoints $( \boldsymbol{k}, \boldsymbol{ \tau })$ are deterministically specified by latent changepoints $( \boldsymbol{k}, \boldsymbol{ \tilde{ \tau } })$ and unknown lags $\boldsymbol{d}$ according to (\ref{eq:latentc}). Therefore, a sample from $( \boldsymbol{k}, \boldsymbol{ \tau })$ can be obtained from a sample from $( \boldsymbol{k}, \boldsymbol{ \tilde{ \tau } }, \boldsymbol{d} )$. 
Next, we propose a sampler from the joint posterior distribution of $( \boldsymbol{k}, \boldsymbol{ \tilde{ \tau } }, \boldsymbol{d} )$, updated from the prior density (\ref{eq:priorjointlatentlags}) by the observed data $\boldsymbol{x}$, thereby providing a means to obtain a sample from the posterior distribution of $( \boldsymbol{k}, \boldsymbol{ \tau })$.

As in Section \ref{sec:binarysampler}, the parameter space is augmented with auxiliary variables $\boldsymbol{u} = (\boldsymbol{u}_1, \ldots, \boldsymbol{u}_T)$ to facilitate the exploration of the state space of the parameters of interest $( \boldsymbol{k}, \boldsymbol{ \tilde{ \tau } }, \boldsymbol{d} )$. Conditionally on latent changepoints $( \boldsymbol{k}, \boldsymbol{\tilde{ \tau } })$ and independently of the lags and the data, for all $t$ and $i < i^\prime$, $u_t(i, i^\prime)$ is assumed to be distributed according to (\ref{eq:condiaux}), such that  $u_t(i, i^\prime) \sim \text{Bernoulli} \left( 1- \exp\{ -\delta \lambda_{i, i^\prime} (1-|\tilde{S}_{i,t} - \tilde{S}_{i^\prime, t} |) \} \right)$, where $\boldsymbol{\tilde{S}}$ is the binary matrix representation of $( \boldsymbol{k}, \boldsymbol{ \tilde{ \tau } })$ according to (\ref{eq:binaryrepresentation}). As described in Section \ref{sec:augmentparam}, it follows that the auxiliary variables 
$\boldsymbol{u}_t$  
%$\boldsymbol{u}_t = (u_{t}(i, i^\prime) )$ 
induce a partition of the time series, $\mathcal{C}_t$, such that, for each cluster $\gamma \in \mathcal{C}_t$, either all time series or no time series in $\gamma$ are affected by a latent changepoint at time $t$.   %As described in 

The joint posterior density of the augmented parameters $( \boldsymbol{k}, \boldsymbol{ \tilde{ \tau } }, \boldsymbol{d}, \boldsymbol{u})$ is 
\begin{align}
\label{eq:jointaugmentedwithlags}
\pi( \boldsymbol{k}, \boldsymbol{ \tilde{ \tau } }, \boldsymbol{d}, \boldsymbol{u} | p, \boldsymbol{\lambda}, \boldsymbol{w}, \boldsymbol{x}, \delta) =  \pi( \boldsymbol{u} | \boldsymbol{\lambda}, \delta, \boldsymbol{k},\boldsymbol{ \tilde{ \tau } }) \pi( \boldsymbol{k}, \boldsymbol{ \tilde{ \tau } }, \boldsymbol{d}  | p, \boldsymbol{\lambda}, \boldsymbol{w}, \boldsymbol{x}).
\end{align}
To sample from the posterior distribution of $( \boldsymbol{k}, \boldsymbol{ \tilde{ \tau } }, \boldsymbol{d}, \boldsymbol{u})$, or $( \boldsymbol{k}, \boldsymbol{ \tilde{ \tau } }, \boldsymbol{d}, \boldsymbol{u}, \boldsymbol{w} )$ if upper bounds for the lags are \textit{a priori} unknown, the MCMC algorithm discussed in Section \ref{sec:binarysampler} is extended as follows: the birth/death and shift moves are adapted to pairs of latent changepoints and lags; and additional moves are introduced for updating the lags. For the lags, note that according to (\ref{eq:dsetlags}), for all $i=1,\ldots, N$ and $j = 1, \ldots, k_i$,  to maintain monotonicity in the changepoints, 
the lag associated to the $j$th changepoint of the $i$th time series must satisfy
\begin{align}
d_{i,j} \in \mathcal{D}_{j}(w_i, k_i,  \boldsymbol{\tilde{\tau}}_i) = \{ \ell \in \mathbb{N} ; \,  d_{-} \leqslant \ell  \leqslant d^{+} \},
\end{align}
where $d_{-} = \max(0, d_{i, j-1} + \tau_{i, j-i} - \tau_{i, j} +1  )$ and $d^{+} = \min( w_{i},  d_{i,j+1} + \tau_{i,j+1}  - \tau_{i, j} -1 )$; and the full conditional probability distribution of $d_{i,j}$ is such that, for all $d_{i,j} \in \mathcal{D}_{j}(w_i, k_i,  \boldsymbol{\tilde{\tau}}_i)$,
\begin{align}
\pi\left( d_{i,j}  \, | \, \boldsymbol{d}_{-(i,j)},  k_i, \boldsymbol{ \tilde{ \tau } }_i, w_i, \boldsymbol{x}_i  \right) &\propto \mathcal{L}_i \left( \tilde{ \tau }_{i, j-1} + d_{i, j-1}, \tilde{ \tau }_{i, j} + d_{i,j} \right) \mathcal{L}_i \left( \tilde{ \tau }_{i, j} + d_{i,j},  \tilde{ \tau }_{i, j+1} + d_{i, j+1} \right)
\label{eq:fullcondlag}
\end{align}
where $\boldsymbol{d}_{-(i,j)} = \{  d_{i^\prime, j^\prime}; \, (i^\prime, j^\prime)  \neq (i, j) \}$ and $\mathcal{L}_i$ is defined in (\ref{eq:likelihood}). % for all $v, w$.

For the extended cluster updating MCMC algorithm, at each iteration of the algorithm, with $(\boldsymbol{k}, \boldsymbol{ \tilde{\tau} }, \boldsymbol{d}, \boldsymbol{u}, \boldsymbol{w})$ denoting the latest particle of the sample chain, one of the following moves is proposed.

\subsubsection*{Extended birth/death move}
The extended birth/death move proposes the birth or death of a cluster of asynchronous changepoints. First, conditionally on the auxiliary variables, latent changepoints $(\boldsymbol{k}^\prime, \boldsymbol{ \tilde{\tau} }^\prime )$ are proposed according to the birth/death move detailed in Section \ref{sec:binarysampler}: for all time series $i\in \gamma \subseteq \{1, \ldots, N\}$, the birth or death of latent changepoint with position $t$ is proposed. Then, updated lags %$\boldsymbol{d}^\prime$ 
are proposed conditional on $(\boldsymbol{k}^\prime, \boldsymbol{ \tilde{\tau} }^\prime )$: If the birth of changepoints is proposed, then, for all time series $i \in \gamma$, there is $j_i$ such that $\tilde{\tau}_{i,j_i} ^\prime= t $, and the lags $\boldsymbol{d}^\prime_{i} = (d_{i,1}, \ldots, d_{i,j_i-1}, d_{i,j_i}^\prime, d_{i,j_i} \ldots, d_{i, k_i} )$ are proposed for the $i$th time series, where $d_{i,j_i}^\prime$ is sampled from the full conditional distribution  (\ref{eq:fullcondlag}); 
otherwise, if the death of changepoints is proposed, then, for all $i \in \gamma$, there is $j_i$ such that $\tilde{\tau}_{i,j_i}= t$, and the lags $\boldsymbol{d}^\prime_{i} = (d_{i,1}, \ldots, d_{i,j_i-1}, d_{i,j_i+1} \ldots, d_{i, k_i} )$ are proposed for the $i$th time series. 

\subsubsection*{Extended shift move}

The extended shift move proposes to shift the positions of a cluster of asynchronous changepoints. First, latent changepoints $(\boldsymbol{k}^\prime, \boldsymbol{ \tilde{\tau} }^\prime )$ and auxiliary variables $\boldsymbol{u}^\prime$ are proposed according to the shift move discussed in Section \ref{sec:binarysampler}: for all time series with index $i\in \gamma$, the position $t^\prime$ is proposed for latent changepoint with position $t$. Then, for all $i \in \gamma$, letting $j_i$ denote the index such that $\tilde{\tau}_{i,j_i}^\prime = t^\prime$, propose $d_{i,j_i}^\prime$ from the full conditional distribution (\ref{eq:fullcondlag}).

\subsubsection*{Update of auxiliary variables}
$\delta$ is sampled from its prior distribution and, conditional on latent changepoints $(\boldsymbol{k}, \boldsymbol{ \tilde{\tau} })$, auxiliary variables are sampled from their full conditional distribution (\ref{eq:condiaux}).

\subsubsection*{Update of lags} %, and update of upper bounds for lags}
A pair $(i, j)$ is uniformly chosen from $ \{ (i, j); \, i = 1, \ldots, N \text{ and } j = 1, \ldots, k_i  \}$, and  the lag $d_{i,j}$ is sampled from the full conditional distribution given in (\ref{eq:fullcondlag}).

\subsubsection*{Update of upper bounds for lags}
If the maximal lags $\boldsymbol{w}$ are \textit{a priori} unknown, for a randomly chosen time series with index $i$ it is proposed to update $w_i$ to $w_i^\prime = w_i + \sigma$ with probability $1/2$, and to update $w_i$ to $w_i^\prime = |w_i - \sigma|$ otherwise, where $\sigma$ is drawn from $\text{Geometric}(\rho)$ for some $0<\rho<1$. Proposing to update $w_i$ requires  proposing updated lags $\boldsymbol{d}_i^\prime = (d_{i,1}^\prime, \ldots, d_{i, k_i}^\prime ) \in \mathcal{D}(w_i^\prime, k_i,  \boldsymbol{\tilde{\tau}}_i) $ for the $i$th time series. For $j=1, \ldots, k_i$, given $w_i^\prime$ and $(d_{i,1}^\prime, \ldots, d_{i,j-1}^\prime , d_{i,j+1}\ldots d_{i, k_i})$, the lag $d_{i,j}^\prime$ is sampled from the full conditional distribution (\ref{eq:fullcondlag}).

\section{Estimating changepoint parameters}
\label{sec:bayesestimate}
To summarise the posterior distribution of changepoint parameters for multiple time series, for each time series $i$, following \citet{Green1995}, one may consider the  posterior marginal distribution of the number of changepoints $k_i$, and the posterior distribution of the changepoint positions $\boldsymbol{\tau}_i$ conditional on $k_i$.  
However, in practice, for each time series $i$, it 
may be necessary 
%will also be of interest
 to report a point estimate $(\hat{k}_i, \boldsymbol{\hat{\tau} }_i)$ for the changepoint parameters $(k_i, \boldsymbol{\tau}_i)$. Following normative Bayesian theory, to define an optimal Bayes estimate for changepoints, we propose a loss function 
that evaluates the quality of estimated changepoints. When assessing the cost associated with the estimate $(\hat{k}_i, \boldsymbol{\hat{\tau} }_i)$ of $(k_i, \boldsymbol{\tau}_i)$, both the number and the positions of changepoints must be taken into account. To address this challenge, we use matchings in graphs, as defined in Definition \ref{defn:matchings0} and Definition \ref{defn:matchings}, to define a loss function $L$ for changepoint estimates in Definition \ref{defn:lossfunction}. 
\begin{defn}{\emph{(Maximum matching in a graph).}} Let $B = (V, E)$ be a graph where $V$ is a vertex set and $E\subseteq V \times V $ is an edge set. A matching $M$ in $B$ is a subset of $E$ such that no two edges in $M$ share a common vertex. A maximum matching in $B$ is a matching that is not a subset of a larger matching in $B$.
\label{defn:matchings0}
\end{defn}
\begin{defn}{\emph{(Minimum weight maximum matching in a graph).}} Let $B = (V, E)$ be a graph with weights $w_{i, j} \geqslant 0$ for all $(i,j)\in E$. A minimum weight maximum matching in $B$ is a maximum matching in $B$ for which the sum of weights of the edges is minimised.
\label{defn:matchings}
\end{defn}
\vspace{-4mm}
When $B$ is a weighted bipartite graph, the Kuhn–Munkres algorithm, also known as the Hungarian algorithm,  \citep{Bondy1976} finds a minimum weight maximum matching in $B$; the time complexity of the algorithm is $\mathcal{O}( \text{card}(E) \text{card}(V) + \text{card}(V)^2 \log \log \text{card}(V))$, where $\text{card}(V)$ and $\text{card}(E)$ denote the cardinality of the vertex set and the cardinality of the edge set of $B$, respectively.
\begin{defn}{\emph{(Loss function $L$ for changepoint estimates).}}
Let $\gamma \geqslant0$. For all $k_i, \hat{k}_i  \geqslant 0$,  $\boldsymbol{\tau}_i = ( \tau_{i, 1}, \ldots, \tau_{i, k_i}) \in \mathcal{T}_{k_i}$ and $\boldsymbol{ \hat{\tau} }_i = ( \hat{\tau}_{i,1}, \ldots, \hat{\tau}_{i, \hat{ k}_i })\in \mathcal{T}_{ \hat{k}_i }$ (\ref{eq:deftau}), let $B_i$ be the weighted complete bipartite graph with vertex sets $V_i = \{0, \ldots, k_i \}$ and  $\hat{V}_i = \{0, \ldots, \hat{k}_i \}$, and weights
\begin{align}
\label{eq:distancecp}
w_{i, j, j^\prime} = \min\{ \gamma,   |\tau_{i, j} - \hat{\tau}_{i, j^\prime} | \} 
\end{align}
for all $j \in V_i$ and $j^\prime \in \hat{V}_i$. Given a minimum weight maximum  matching $M_i$ in $B_i$,  
for all $j \in V_i$ and $j^\prime \in \hat{V}_i$, let $m_{i, j, j^\prime} = 1$ if $j$ and $j^\prime$ are matched, that is $( j, j^\prime ) \in M_i$, and $m_{i, j, j^\prime} =0$ otherwise. Then, define the loss to be 
\begin{align}
\label{eq:lossfunction}
L \left[ (\hat{k}_i, \boldsymbol{\hat{\tau}_i } ), (k_i, \boldsymbol{\tau}_i) \right] = \gamma |\hat{k}_i - k_i | + \frac{1}{2}\sum_{j \in V_i} \sum_{j^\prime \in \hat{V}_i } m_{i, j, j^\prime}w_{i, j, j^\prime }.
\end{align}
\label{defn:lossfunction}
\end{defn}
\vspace{-2mm}
Consider the complete bipartite graph $B_i$ with independent vertex sets $V_i = \{0, \ldots, k_i \}$, $\hat{V}_i = \{0, \ldots, \hat{k}_i \}$ and weights (\ref{eq:distancecp}). A minimum weight maximum matching $M_i$ in $B_i$ gives a matching of the elements of $\boldsymbol{\hat{\tau} }_i$ and $\boldsymbol{\tau}_i$ that minimises the sum of distances (\ref{eq:distancecp}) between matched changepoints. Given $M_i$, according to the loss function (\ref{eq:lossfunction}), the cost associated with the estimate $(\hat{k}_i, \boldsymbol{\hat{\tau} }_i)$ of $(k_i, \boldsymbol{\tau}_i)$ is then obtained by adding the cost $\gamma$ for each unmatched changepoint and the total distance between matched changepoints. Note that according to (\ref{eq:distancecp}), the cost of matching two changepoint positions, that are separated by more than $\gamma$ time units, is equal to the cost of an unmatched changepoint, namely $\gamma$. Therefore, the loss function $L$ takes into account both the number and the positions of changepoints, and the cost $\gamma$ is chosen to be the maximum acceptable distance between a changepoint position and its estimated position. 
The optimal Bayes estimate $(\hat{k}_i, \boldsymbol{\hat{\tau} }_i)$ is the changepoint parameters that minimise the expected posterior loss with respect to the posterior marginal distribution of the changepoints $(k_i, \boldsymbol{\tau}_i)$. 

%Partially as a consequence of the intractability of the normalising constant of the posterior density of changepoint parameters, inference via simulations is required; Section \ref{sec:inferenceMCMC} discusses algorithms to obtain a sample of approximate draws from the posterior distribution of changepoints for all time series. 
Given an approximate sample from the posterior distribution (Section \ref{sec:inferenceMCMC}),  an approximate Bayes estimate $(\hat{k}_i, \boldsymbol{ \hat{\tau} }_i )$ for each series can be  identified numerically by finding within the sample the changepoint parameters  that minimise the estimated posterior expected loss.

\textcolor{corrections}{Appendix \ref{sec:appendixsimustudy} %in the supplementary material \citep{SuppCPGTS} 
presents a simulation study that demonstrates the model introduced in Section \ref{sec:newprior} and the MCMC sampling strategy discussed in Section \ref{sec:inferenceMCMC}, using the loss function introduced in this section. In particular, various graphs and changepoint parameters are considered to illustrate the flexibility of the proposed model, and the convergence of the sampler is demonstrated under a wide range of settings.}

\section{Red team detection in network authentication data from LANL}

\label{sec:cyberapplication}
This section presents results of an analysis of the LANL network authentication data presented in Section \ref{sec:motivationalsection} that demonstrates the utility of the graphical changepoint model proposed in Section \ref{sec:newprior}. %, through a comparison with the standard model for independent changepoints across time series (\ref{eq:classicprior}).

\vspace{-2mm}

\subsection{Presence of a red team}
The occurrence of a red team exercise during the first month of the data collection provides surrogate intruder behaviour in the authentication data \citep{akent-2015-enterprise-data}.  In particular, $103$ user IDs are known to have been used by the red team. We show the graphical model for dependent changepoints can combine evidence from multiple users which are linked in the network, to detect chains of quasi-synchronous weak signals for changes in the authentication activity of red team users, whilst limiting the number of false alerts.

\begin{figure}[t!]
% \vspace{-3mm}
      \centering
      \includegraphics[width=0.80\textwidth]{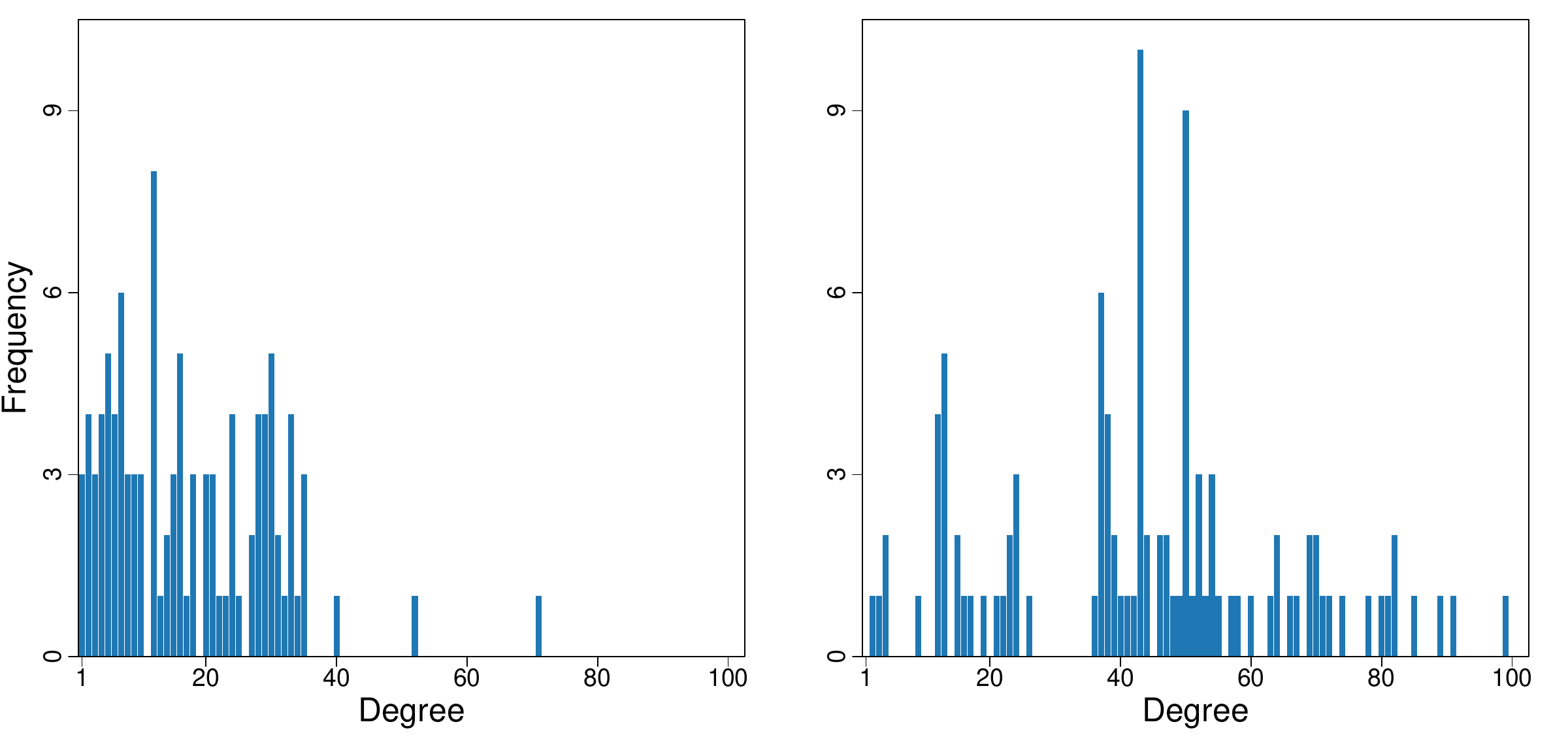}
    \caption{Degree distribution of users in the LANL network represented by the graph $G^\prime$. Left panel: counts for legitimate users. Right panel: counts for red team users. 
    % and for legitimate users $i \in N $ are displayed in the left and right panel, respectively. %indicated in red and in blue, respectively.
    }
     \label{fig:kentdegree}
     %\vspace{-3mm}
\end{figure}

For our demonstration purposes, it suffices to examine %the authentication activity of 
a subset of the full LANL network of users, which is represented by the graph $G = (V, E)$ defined in Section \ref{sec:graphusers}. Let $R$ denote the set of red team users and let $B \subseteq \{ u \in V \setminus R : \, \exists r \in R \, \text{ s.t. } (u, r) \in E  \}  $ denote $\mathrm{card}(R)$ randomly selected users that are not labelled as red team users in the data but are linked to red team users on the network. 
The focus is on the network corresponding to the subgraph $G^\prime$ induced in $G$ by the set of users $V^\prime = R \cup B$. Figure \ref{fig:kentdegree} shows the degree distribution of the $206$ users in $G^\prime$. Red team users tend to have a greater degree in $G^\prime$ than legitimate users; to traverse the network towards high value targets, intruders tend to take control of users that are highly linked on the network. 
%. To traverse the network towards high value targets, intruders tend to take control of users that are highly linked on the network. 

\subsection{Changepoint modelling}
Recall from Section \ref{sec:motivationalsection} that, for each user  $i \in V^\prime$, the data $x_{i, 0}, \ldots, x_{i, T}$ consist of hourly counts of network logons per source computer for the first month of data collection as defined in (\ref{eq:datacyber}), which are now assumed to follow the model specified in (\ref{eq:classicconiid}) for multinomial data. 
Different graphical changepoint priors are considered to demonstrate the benefits of encoding prior beliefs about cyber-attacks. To encode prior belief that signals for changes resulting from an attack are likely to occur at similar times across users that are linked in the network $G^\prime$, the graphical changepoint prior specified in (\ref{eq:priorextended}) is considered with an identical edge weight parameter $\lambda > 0$, as defined in Section \ref{sec:aspecialcase}, for all pairs of time series corresponding to users that are linked in $G^\prime$. Moreover, for comparison purposes, the graphical changepoint prior %for dependent changepoints specified in
 (\ref{eq:priorextended}) is also considered assuming the complete graph defined in Section \ref{sec:completegraph} such that $\lambda_{i, j} = \lambda > 0$ for all pairs of users $(i, j) \in V^\prime \times V^\prime$. %Note that in each case, fixing 
With $\lambda = 0$, the two graphical changepoint priors of interest correspond to the standard changepoint model assuming independence of changepoints across time series (\ref{eq:standard0}). 

For comparison purposes and to illustrate the flexibility of the proposed model, a collection of changepoint prior parameters are considered: 
$\bar{p} \in \{-30, -50, -70\}$ and $\lambda = \lambda_s |\bar{p}| / n$ with $\lambda_s \in \{0, 0.5, 0.6, 0.7, 0.8\}$, where $n$ denotes the average node degree in the graph. %$G^\prime$. 
 Moreover, different assumptions for the upper bounds $\boldsymbol{w}$ for the lags are compared: the \emph{zero window} assumption with $w_i=0$ for all $i$, implying signals for attacks are assumed to be synchronous across users; 
 and, the \emph{variable window} assumption with $w_i \sim \text{Geometric}(0.9)$ for all $i$, admitting signals for attacks may be asynchronous across users. 
 
 For each simulation, a sample of size $1 \, 000 \, 000$ was obtained from the posterior distribution of the changepoints via the MCMC algorithm proposed in Section \ref{sec:mcmcwithlags}, with a burn-in of $300 \, 000$ iterations; the Bayes estimate for changepoints corresponding to the loss function (\ref{eq:lossfunction}) with $\gamma = 48$ was then derived from the sample.

\subsection{Results}

\begin{figure}[b!]
\vspace{-2mm}
\centering
        \includegraphics[width=0.86\textwidth]{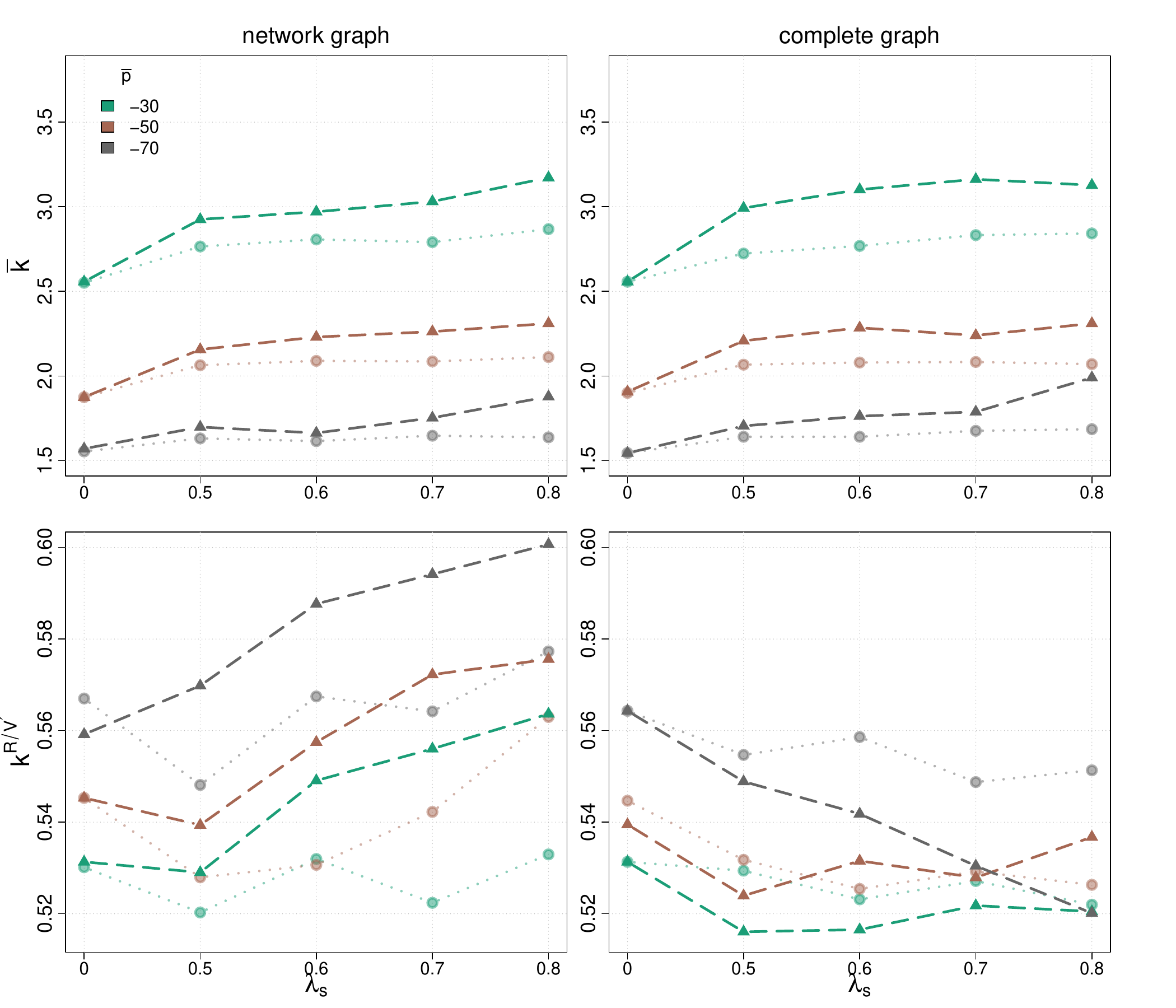}
    \caption{Average number of changepoints  per time series $\bar{k}$ (top row), and proportion of changepoints impacting redteam users $k^{R / V^\prime}$ (bottom row), assuming the network graph (left column) and the complete graph (right column), 
    for different assumptions for the upper bounds for the lags - \textit{zero window} (circles) and \textit{variable window} (triangles), and for a collection of  prior parameters $\bar{p}$ (identified by distinct colours) and  $\lambda_s$.}
     \label{fig:kentcpnuum}
\end{figure}

\begin{figure}[b!]
\centering
        \includegraphics[width=0.86\textwidth]{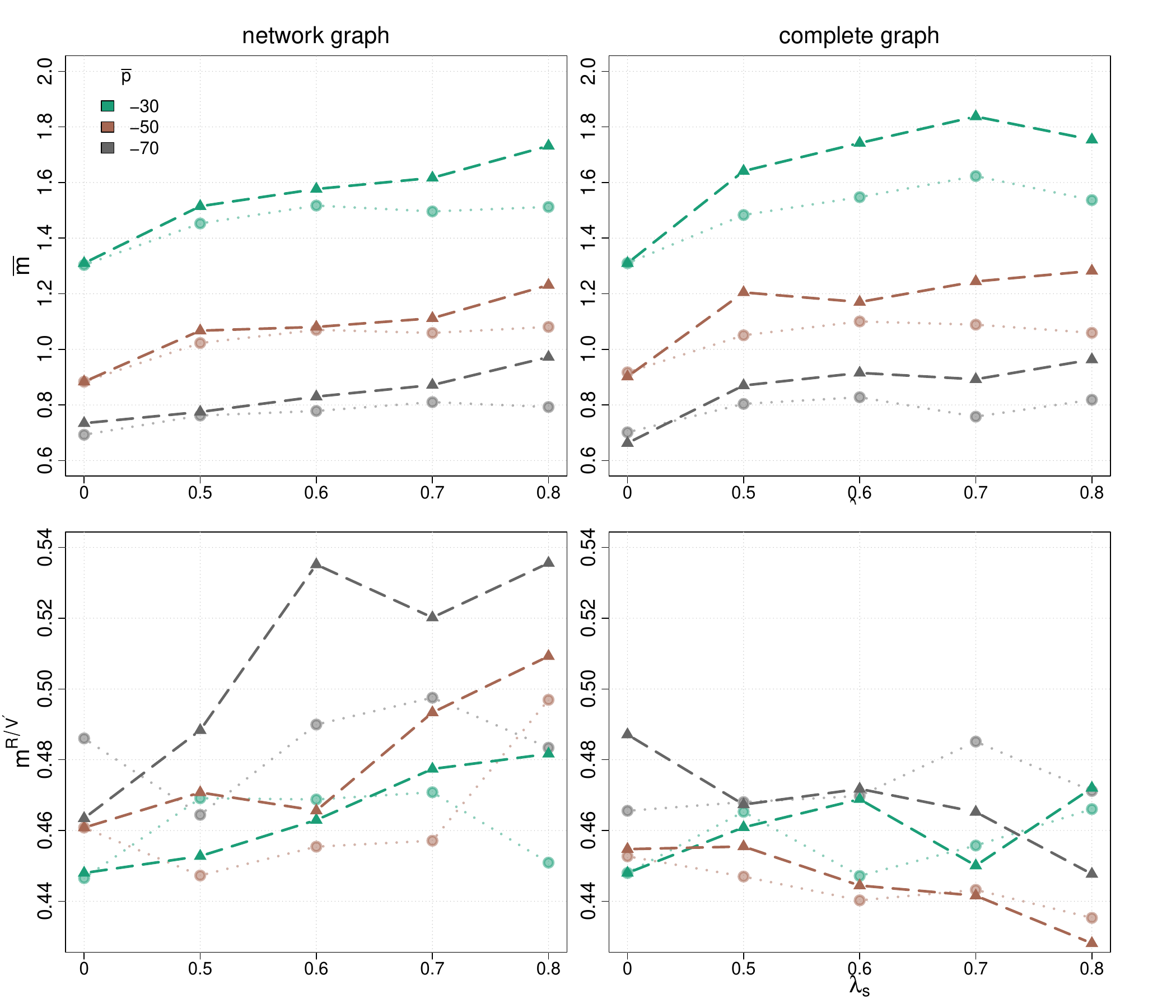}
    \caption{Estimated $\bar{m}$ and $m^{R / V^\prime}$ (\ref{eq:minew}), assuming the network graph (left column) and the complete graph (right column), 
    for different assumptions for the upper bounds for the lags - \textit{zero window} (circles) and \textit{variable window} (triangles), and for a collection of prior parameters $\bar{p}$ (identified by distinct colours) and  $\lambda_s$.}
      \label{fig:kentcpnuum2}
      \vspace{-2mm}
\end{figure}

For each user in the network, each estimated changepoint represents a piece of evidence for possible malicious behaviour that might require further investigation by cyber analysts. Identifying inferred changepoints for time series corresponding to legitimate users $i \in B$ as false alerts, it is meaningful to compare models in terms of the estimated number of changepoints per time series $i\in V^\prime$, 
\begin{align}
\bar{k} = \frac{1}{\mathrm{card}(V^\prime)} \sum_{i \in V^\prime} k_i,
\end{align}
and the proportion of estimated changepoints that impact redteam users,
\begin{align}
k^{R / V^\prime} = \frac{\sum_{i \in R} k_i}{\sum_{i \in V^\prime} k_i} . 
\end{align}
Moreover, since cyber-attacks tend to be identified through clusters of behavioural changes across machines that are linked on the network, it is of interest to prioritise for investigation the estimated changepoints that belong to clusters of quasi-synchronous changepoints on the network.
Given some time window $\varpi \geqslant 0$, 
let the weight of $\tau_{i, j}$ be 
\begin{align}
c_{i, j}^{\varpi} = \frac{ n_{i, j}^{\varpi} + 1 }{ n_{i} + 1},
\end{align}
where 
\begin{align}
 n_{i, j}^{\varpi} = \mathrm{card}( \{ (i, i^\prime) \in E :  \;  \exists j^\prime \in \{1, \ldots, k_{i^\prime} \} \text{ s.t. } |\tau_{i, j} - \tau_{i^\prime,  j^\prime}| \leqslant \varpi \})
\end{align}
denote the number of users linked to user $i$ in $G^\prime$ that are impacted by a changepoint within $\varpi$ hours of $\tau_{i, j}$, and where $n_{i} = \mathrm{card}( \{ (i, i^\prime) \in E \} )$ is the degree of node $i$ in $G^\prime$, such that $( n_i + 1  )^{-1}  \leqslant c_{i, j}  \leqslant 1$. The larger the weight $c_{i, j}$, the more connected $\tau_{i, j}$ to other changepoints across the network. To take into account both the number of changepoints and %
their connectedness, 
for each user $i$, changepoint estimates are also compared via the sum of weights %statistic
\begin{align}
\label{eq:mi}
m_{i} \equiv m_{i}^{\varpi} = \sum_{j = 1}^{k_i} c_{i, j}^{\varpi},
\end{align} 
such that $0 \leqslant m_i  \leqslant k_i$. Note that for each user $i$, $m_i$ increases 
with both the number of changepoints and their weights. Let 
\begin{align}
\label{eq:minew}
\bar{m} = \frac{1}{\mathrm{card}(V^\prime)} \sum_{i \in V^\prime} m_i, \quad m^{R / V^\prime} = \frac{\sum_{i \in R} m_i}{\sum_{i \in V^\prime} m_i}
\end{align}
be the average sum of changepoint weights per user and the proportion of changepoint weights associated to redteam users, respectively.

For each choice of graph and changepoint prior parameters, Figure \ref{fig:kentcpnuum}
displays the estimated values of $\bar{k}$ and $k^{R/V^{\prime}}$, and Figure \ref{fig:kentcpnuum2} displays the estimated values of $\bar{m}$ and $m^{R/V^{\prime}}$ assuming $\varpi = 48$. 
As $\bar{p}$ increases, weaker evidence is required to infer changepoints, and therefore, for each graph, $\bar{k}$ and $\bar{m}$ increase. % for both red team users and legitimate users.
As $\lambda$ increases, the estimates for $\bar{k}$ and $\bar{m}$ tend to increase for each graph, but the estimates for $k^{R/V^{\prime}}$ and $m^{R/V^{\prime}}$ tend to increase only when assuming the network graph. 
This follows because the graphical changepoint model assuming the network graph successfully encodes prior knowledge that cyber attacks tend to correspond to coordinated activity across multiple users linked by network connectivity, and consequently, as $\lambda$ increases, it detects weaker signals for behavioural changes that correspond to red team activity, whilst crucially limiting the number of false alerts. For the proposed model with the complete graph, all time series are connected, and therefore, as $\lambda$ increases, weaker signals for changes are detected for red team activity but also for legitimate activity, which would impede fast identification of the attack.

Moreover, results show the benefits of the model extension which relaxes the assumption that signals for attacks are synchronous across users. As $\lambda$ increases, when assuming the network graph,  the increase of the estimates for $k^{R/V^{\prime}}$ and $m^{R/V^{\prime}}$ tend to be greater for the \emph{variable window} scenario than for the \emph{zero window} scenario. In contrast with the \emph{zero window} scenario, the \emph{variable window} scenario admits attacks may result in quasi-synchronous behavioural changes across the network, and consequently clusters of nearby but not necessarily synchronous weak signals for changepoints are detected across red team users. 

The results show that, in comparison with the standard model for independent changepoints across time series, the proposed graphical changepoint model provides a flexible tool for cyber-analysts to incorporate expert knowledge in changepoint analysis for network monitoring, thereby facilitating network intrusion detection.

 \vspace{-4mm}

\section{Discussion}
\label{sec:discuss}
This article considers a setting with $N$ time series (\ref{eq:introdata}) subject to changepoints, where it is desirable to encode in the changepoint prior, by means of a graph $G= (V, E)$ on $N$ nodes corresponding to each of the time series, that pairs of time series $(i, i^\prime) \in E$ are \textit{a priori} more likely to be impacted by simultaneous changepoints. % 
This setting is adapted to the application in cyber-security  where  
each node in $V$ corresponds to a time series representing the authentication activity of a network user, and an edge $(i, i^\prime) \in E$ indicates that it is believed \textit{a priori} that attackers may switch credentials between user $i$ and user $i^\prime$ at any time of the data collection period, so that users $i$ and $i^\prime$ are \textit{a priori} more likely to be impacted by quasi-simultaneous behavioural changes.

However, for some applications, it might be restrictive to assume that prior beliefs on which time series are likely to be impacted by simultaneous changepoints do not vary over time. 
For example, consider the following application in cyber security. Using system log data, it can be of interest to monitor the process activity of computers, which may be subject to changes when attackers perform malicious activity such as the installation or the execution of  malware. Moreover, attackers will typically need to communicate with  compromised computers to simultaneously execute malicious commands on these computers. As a result, the process activity of computers $i$ and $i^\prime$ are more likely to be subject to simultaneous changes when some source computer simultaneously communicates to both $i$ and $i^\prime$. 
For such a setting, it would be more suitable to specify a time series of graphs $\{ G_t = (V, E_t) \, | \, E_t  \subseteq  V \times V , t \geqslant 0\}$, such that pairs of time series $(i, i^\prime) \in E_t$ are \textit{a priori} more likely to be impacted by simultaneous changepoints at time $t$. Each node in $V$ would correspond to a time series representing the process activity of a computer in the network, and an edge $(i, i^\prime)\in E_t$ would indicate that  communication events occurred at time $t$ from some source computer to both $i$ and $i^\prime$, so that computers $i$ and $i^\prime$ are \textit{a priori} more likely to be impacted by simultaneous behavioural changes at time $t$.
For networks where many computers may leave or enter during the data collection period, a further model extension could consider relaxing the assumption that $V$ is fixed, 
specifying a time series of graphs $\{ G_t = (V_t, E_t) \, | \, E_t  \subseteq  V_t \times V_t , V_t \subset \mathbb{N},  t \geqslant 0\}$ such that $V_t$ is the node set of computers active in the network at time $t$.   
With the introduction of time-dependent edge weight parameters $\boldsymbol{\lambda} = (\lambda_{i, i^\prime, t})$ such that $\lambda_{i, i^\prime, t} >0$ if and only if $(i, i^\prime) \in E_t$, these model extensions would present no theoretical complication, with a straightforward   adaption of the graphical changepoint prior and the proposed sampling strategy.

\section*{Supplementary material}
%Thepythoncode and datasets are available athttps://www.github.com/fraspass/sbm.
The \textit{python} code and the data are available at \url{https://github.com/karl-hallgren/cp_on_graph_of_timeseries/}

\bibliography{bibli.bib}

\section*{Acknowledgements}
The authors thank Niall Adams for stimulating discussions about this work. The authors acknowledge funding from EPSRC. Research presented in this article was supported by the Laboratory Directed Research and Development program of Los Alamos National Laboratory (New Mexico, USA) under project number 20180607ECR and Los Alamos National Laboratory.

\appendix

\clearpage

\section*{Appendices}

\section{Motivational example: changepoint detection in cyber-security}
\label{sec:appendixcyber}
This section gives further details on the motivational example discussed in Section \ref{sec:motivationalsection}.
\subsection{Data processing}
\label{sec:appendixcyberdata}
%The motivation example discussed in Section \ref{sec:motivationalsection}.   
With the aim of modelling network activity that is human driven rather than high frequency and automated, the network authentication data, which are available online at \url{https://csr.lanl.gov/data/cyber1}, were filtered as follows: events involving the same pair of user and source computer more than $5 \, 000$ times were removed; computers that act as source computers for more than $300$ distinct users were discarded. The resulting data involves $11 \, 985$ distinct users and $12 \, 947$ distinct source computers.

\subsection{Limitations of changepoint independence across time series}
\label{sec:appendixcyberstudy}
\begin{figure}[h!]
\centering
        \includegraphics[width=0.95\textwidth]{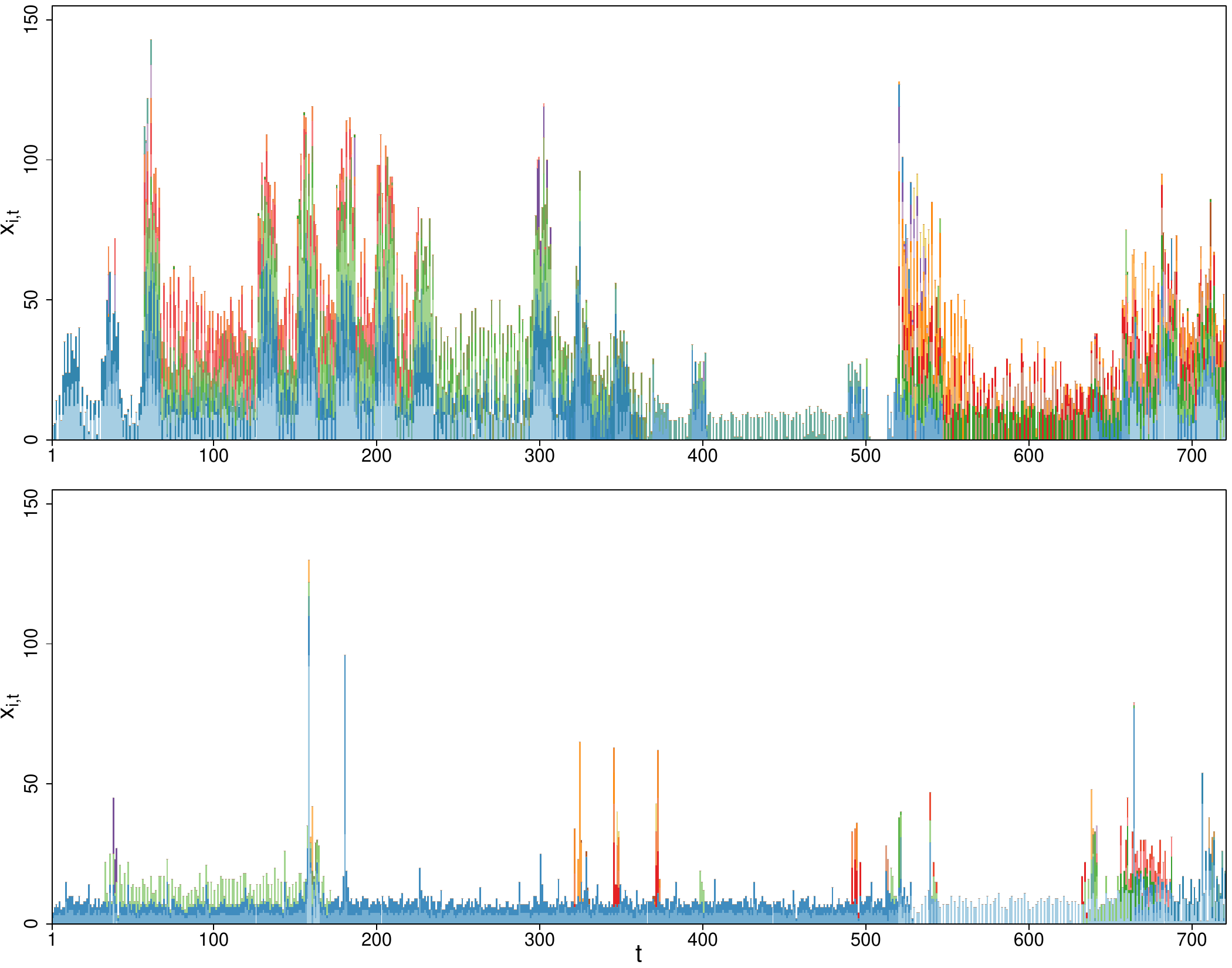}
    \caption{Hourly counts of network logons per source computer for two users over the first month of data collection. Source computers are identified by distinct colours. Top panel corresponds to user U342@DOM1, and bottom panel to user U86@DOM1. 
    }
    \label{fig:motiv1}
\end{figure}

\begin{figure}[b!]
\centering
        \includegraphics[width=0.95\textwidth]{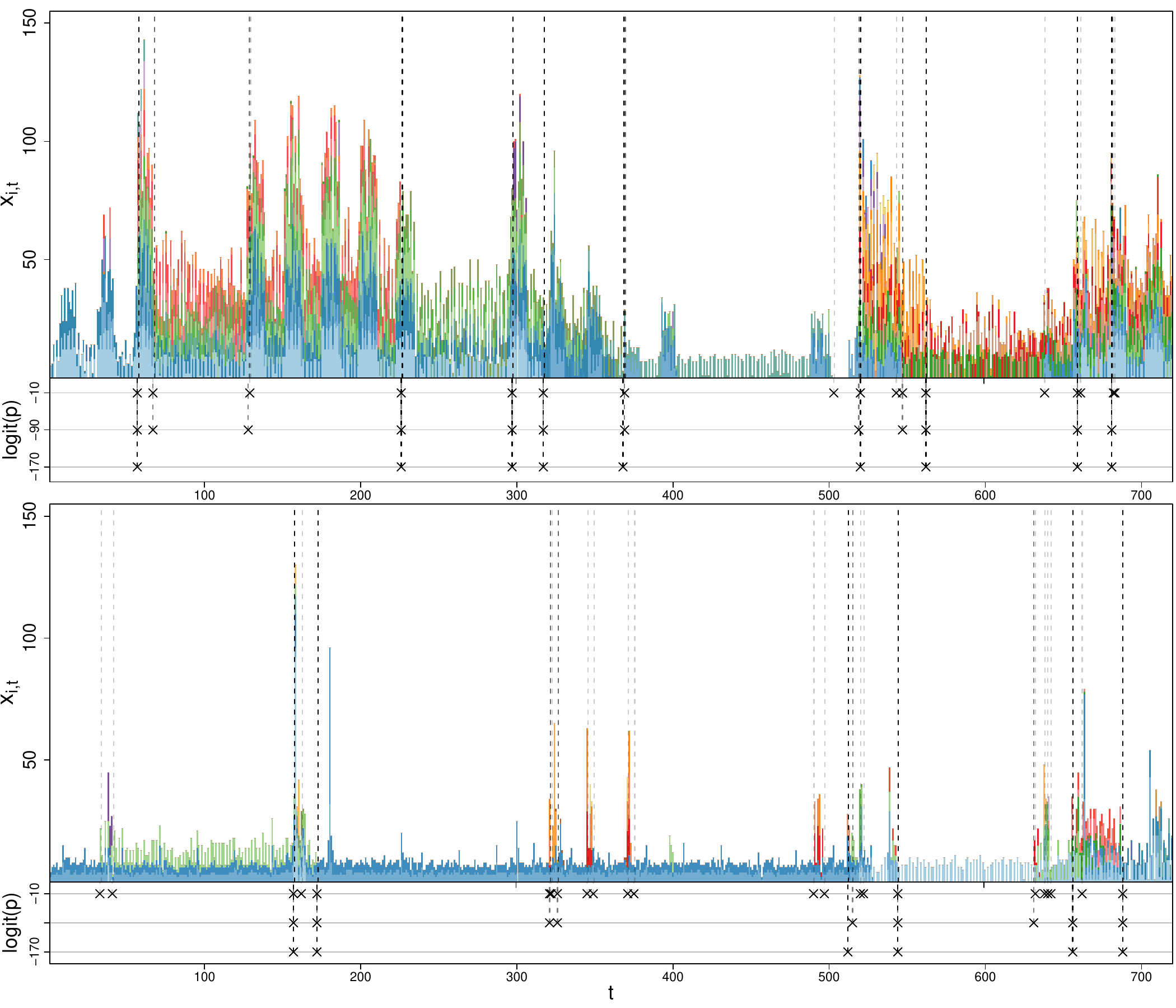}
    \caption{Hourly counts of network logons per source computer  for two users over the first month of data collection; top panel corresponds to user U342@DOM1 and bottom panel to user U86@DOM1. Crosses indicate positions of Bayes changepoint estimates, according to the loss function $L$ with $\gamma=40$, corresponding to three different Bernoulli parameters $p$.}
    \label{fig:motiv2}
\end{figure}

To detect occurrences of malicious activity in the network, the authentication activity of each user is monitored via hourly counts of network logons per source computer. Figure \ref{fig:motiv1} displays the first month of data for two distinct users that are linked on the network. The data are assumed to follow the changepoint model specified in Section \ref{sec:motivationalsection}. Assuming further that changepoints are independent with prior distribution given in (\ref{eq:classicprior}), for a selection of Bernoulli parameters $p$, a sample from the posterior distribution of changepoints was obtained via the reversible jump MCMC algorithm proposed in \citetsupp{Green1995}, adapted to discrete time changepoints. 
In Figure \ref{fig:motiv2}, for different Bernoulli parameters, the crosses and the red lines indicate the positions of Bayes changepoint estimates corresponding to the loss function $L$ given in Definition \ref{defn:lossfunction} with $\gamma = 40$; it is noticeable that $p$ controls \textit{a priori} the level of granularity of the segmentation of the data. The smaller $p$, the stronger the evidence required from the data to suggest a posteriori inferred changes. No choice of $p$ seems satisfactory: choosing a small value for $p$ will limit the number of false alerts due to noise and user specific legitimate activity; yet it will also prevent the detection of weak signals for changes shared by different users which are linked in the network, that may be of great interest.%

\vspace{5mm}

\clearpage

\section{Proof of Proposition \ref{prop:cardinalityd}}
\label{sec:appendixproof}
\begin{itemize}
\item[(i)] Let $w \geqslant 0$, $k  \geqslant 0$ and $ \tau_{1:k} = (\tau_1, \ldots, \tau_k ) \in \mathcal{T}_k$. We introduce some notations to faciliate the  proof. For all $0 \leqslant  j \leqslant l \leqslant k$,  let 
\begin{align*}
\Omega_{j:l}  = \Omega_{j:l} (w, k, \tau_{1:k} ) &=   \{ d_{j:l} = (d_j, \ldots,  d_{l})  : \forall i, \, \, 0 \leqslant d_i < \rho(i, 0) = \min\{w+1, T+1 - \tau_i \} \} 
\end{align*}
and
\begin{align*}
\mathcal{Q}_{j:l} \equiv \mathcal{Q}_{j:l}(w, k, \tau_{1:k} )  = \{  d_{j:l} = (d_j, \ldots,  d_{l}) \in \Omega_{j:l}  :  \tau_j + d_j \geqslant \tau_{j+1} + d_{j+1} \geqslant \cdots   \geqslant \tau_{l} + d_{l} \}. 
\end{align*}
Note that $\mathrm{card}(\Omega_{j:l}) = \prod_{i=j}^{l} \rho(i, 0)$ and $\mathrm{card}(\mathcal{Q}_{j:l}) = Q(j, l-j)$ where $Q$ is defined in (\ref{eq:defQ}). 

It is immediate that if $k=0$ then (\ref{eq:Dcard}) holds. If $k=1$ then $\mathrm{card}(\mathcal{D}(w, k, \tau_{1:k} ) ) = \min\{w + 1,T+1 - \tau_1 \}  = Z(1)$. We show by induction that (\ref{eq:Dcard}) holds for $k>1$. If $k=2$ then 
\begin{align}
\mathrm{card}( \mathcal{D}(w, k, \tau_{1:k} ) ) &= \mathrm{card}( \Omega_{1:2} ) - \mathrm{card}(\mathcal{Q}_{1:2} ) \nonumber \\ \nonumber
& = \rho(1, 0)\rho(2, 0) - Q(1, 1) \\  \nonumber
&= Z(1) Q(2, 0) - Z(0)Q(1, 1) \\  \nonumber
&= Z(2).
\end{align}
% \rho(1, 1)( \rho(1, 1) + 1 )/2 \\

Assume that  (\ref{eq:Dcard}) holds for $k=1, \ldots, k^\prime-1$ for some $k^\prime > 2$. We show that (\ref{eq:Dcard}) holds for $k = k^\prime$. We have
\begin{align}
\mathrm{card}( \mathcal{D}(w, k, \tau_{1:k} ) ) & = \sum_{j=1}^{k} (-1)^{k-j}  \mathrm{card}( \{  d_{1:k} \in \Omega_{1:k} :  d_{1:(j-1)} \in \mathcal{D}(w, j-1, \tau_{1:(j-1)} ) , \, d_{j:k} \in  \mathcal{Q}_{j:k}   \} )   \nonumber \\
& = \sum_{j=1}^{k} (-1)^{k-j}  \mathrm{card}( \mathcal{D}(w, k-j, \tau_{1:(k-j)} ) ) \,  \mathrm{card}( \mathcal{Q}_{j:k}  )   \nonumber \\
& =  \sum_{j = 1}^{k}  (-1)^{k-j}  Z(j-1)Q( j, k -j )   \nonumber \\
& = Z(k).   \nonumber
\end{align}

\item[(ii)] Follows immediately from (\ref{eq:dsetlags}). %   . Note that from (i). 
\end{itemize}

\clearpage

\section{Markov chain Monte Carlo inference}
\label{sec:appendixMCMCalgo}

\textcolor{corrections}{Section \ref{sec:inferenceMCMC} proposed an auxiliary variable MCMC algorithm to sample from the graphical changepoint model. This section gives some indications on the time complexity of the algorithm and discusses possible extensions for settings where segment parameters cannot be marginalised. }

\subsection{Time complexity and initialisation of the sampler}
\label{sec:timecomplexity}
The move proposing to update auxiliary variables $\boldsymbol{u} = (\boldsymbol{u}_t)$ from their full conditional distribution (\ref{eq:condiaux}) is independent of preceding auxiliary variables in the sample chain and is accepted with probability $1$.      Moreover, the auxiliary variables $\boldsymbol{u}$ provide a means to explore the changepoint parameter space but it is not of interest to store a sample from the posterior distribution of $\boldsymbol{u}$. 
% The move proposing to update auxiliary variables $\boldsymbol{u} = (\boldsymbol{u}_t)$ from their full conditional distribution (\ref{eq:condiaux}) is accepted with probability $1$.
Therefore, for all $t$, the sampling of the auxiliary parameters $\boldsymbol{u}_t$ is performed only when a subsequent move depends upon $\boldsymbol{u}_t$, for example when a move proposes to update a cluster of changepoints at time $t$. As a result, the cost of updating auxiliary variables is at most $\mathcal{O}(\mathrm{card}(E))$ per iteration.
Moreover, for the shift and the birth/death moves that propose to update changepoints for a cluster of time series $C$, 
an $\mathcal{O}(T)$ Bayes factor calculation for the model in (\ref{eq:classicconiid}) is required for each time series in $C$. 
% it is required to compute a Bayes factor, with computational cost $\mathcal{O}(T)$ for the model in (\ref{eq:classicconiid}). 
If $\delta=0$ then $\mathrm{card}(C)=1$; otherwise, the expected size of the cluster $1 \leqslant \mathrm{card}(C) \leqslant N$ will tend to increase with $N$. Hence, the computational cost of moves proposing to update changepoints is at most $\mathcal{O}(TN)$. 

To speed-up the convergence of the sampler, we may consider the following initialisation of the sample chain: latent changepoints are set to be the optimal Bayesian estimates, according to the loss function later specified in Definion \ref{defn:lossfunction},  corresponding to the standard model with  independent changepoints across time series with prior $\pi(\boldsymbol{k}, \boldsymbol{\tau } | p )$,  obtained in parallel for each time series; initial lags, upper bounds for lags and auxiliary variables are set to $0$. As a result, the burn-in for the sampler begins with a sensible positioning of changepoints at a limited cost: no iterations for the burn-in of the joint sampling of changepoints across time series are wasted identifying strong signals for changepoints, relative to $\bar{p}$. 

Section \ref{sec:mcmcdiagnosticruntime} presents results of a simulation study demonstrating the feasibility of the proposed inference via MCMC for a range of $N$ and $T$ that can correspond to realistic scenarios. For high-dimensional applications where the time until reaching suitable convergence can be prohibitively long, future work could examine analytical approximations to the posterior distribution of changepoints, for example via variational methods \citepsupp{bleivaria}.

\subsection{Possible sampling algorithm extensions for segment parameters}
\label{sec:possiblemodelextension}
The proposed sampling strategy assumes that segment parameters can be marginalised to compute the conditional likelihood of the data given changepoints, $\mathcal{L}( \boldsymbol{x} | \boldsymbol{k}, \boldsymbol{\tau})$. Although this assumption is appropriate in many applications, as discussed in Section \ref{sec:generalmodel}, it will not always be the case; %more sophisticated models might be required in some situations, 
for example when segment parameters are dependent across segments  \citepsupp{CHIB1998, CHIB2019, Fearnhead2011EfficientBA}. If segment parameters cannot be marginalised, the proposed reversible jump MCMC algorithm can be adapted to sample segment parameters alongside changepoint parameters as in \citetsupp{Green1995}: a move is introduced to sample segment parameters conditional on changepoints; the moves proposing updated changepoint parameters are extended to propose suitable parameters for the segments that are affected by the proposed change of changepoint parameters; note that the proposal distributions for segment parameters depend on the segment model of interest.

Moreover, an interesting model extension of the changepoint model in (\ref{eq:classicconiid}) %where segment parameters are dependent, but can be marginalised,
consists in specifying that segment parameters may be shared across segments: each segment parameter $\theta_{i, j}$ can take one of a finite, but unknown, number of states. Segments that share the same parameter, possibly across time series, are said to be under the same regime, and regime parameters can still be marginalised for conjugate probability models \citepsupp{Bolton2018}. 
In this setting, the proposed MCMC algorithm would need to be adapted to sample the changepoint paramaters, the number of regime and a regime for each segment % number of regimes and a regime for each segment alongside the changepoint parameters 
following \citetsupp{Bolton2018}.

\section{Simulation study}

% \vspace{-2mm}
\label{sec:appendixsimustudy}
\textcolor{corrections}{This section presents a simulation study to demonstrate the model for dependent changepoints proposed in Section \ref{sec:newprior} and the sampling strategy discussed in Section \ref{sec:inferenceMCMC}.} 
In particular, various graphs and changepoint parameters are considered to illustrate the flexibility of the proposed model. 
Moreover, the proposed model is compared with the standard model for independent changepoints, which assumes all edge weight parameters are null (\ref{eq:standard0}), and 
with MVCAPA \citepsupp{fisch}, which can borrow strength across multiple time series to detect synchronous changepoints across a subset of the time series
but assumes \textit{a priori} changepoint locations are exchangeable across time series.

\subsection{Synthetic data}
Synthetic data for the simulation study were sampled according to the changepoint model for $N$ time series each of length $T$ defined in (\ref{eq:classicconiid}) with, for all $i$, $f_i( \, \cdot \, | \,  \theta_{i, j})$ corresponding to $\text{Poisson}(\theta_{i, j})$, for a range of values of $N \in \{30, 130, 230, 330\}$ and $T\in \{300, 3 \, 300, 6 \, 300, 9 \, 300\}$.

For simulating data, let $C\subset\{1,\ldots,N\}$ be a non-empty subset of the time series indices, and let $\bar{C}$ be the set complement such that $C\cup\bar{C}=\{1,\ldots, N\}$. 
Then let $k_i = k$ if $i \in C$, where $k$ is uniformly sampled from $\{1, \ldots, 7\}$, and $k_i=0$ otherwise, such that $C$ denotes those series which experience changepoints while those in $\bar{C}$ are not impacted by changepoints.
The impact of encoding prior information on the dependence structure of changepoints across time series on a graph will be investigated by means of further divisions of the set $C$ into subsets. 
We fixed $C = C_1 \cup C_2 \cup C_3$ for some disjoint sets $C_1= \{1, 2, \ldots,  \floor*{N/5} \}$, $C_2 = \{  N, N-1\} \cup \{N-\floor*{N/5}, N-\floor*{N/5}-1 \} \cup \{N-2\floor*{N/5}, N-2\floor*{N/5}-1 \}$ and $C_3=\{\floor*{N/2} -1\}$. 
As illustrated in Figure \ref{fig:graphsimul} for $N=30$ time series, according to both the $6 \times 5$ lattice graph or the $2$-chain graph, the elements of $C_1$ and $C_2$ cluster according to the graph, whereas the single element of $C_3$ has no neighbours in $C$ so that it is isolated in the graph. 
Moreover, different scenarios for  changepoint positions were considered: for all $i \in C$, we set $\tau_{i, j} = \floor*{jT/(k_i+1)}  + v_{i, j}$, where $v_{i, j}$ is sampled from $\{-v, 0, v\}$, for increasing levels of asynchrony for the changepoints $v \in \{0, 5, 10, 15\}$. For the scenario with $v=0$, changepoints are synchronous. % for the set-up with $v=0$. 
Finally, for all $i \in C$, the signal strength for changepoints may be quantified by 
\begin{align}
\mu_i = \frac{1}{k_i} \sum_{j=1}^{k_i} \log  \left\lbrace E \left( \frac{\mathcal{L}_i(\tau_{j-1}, \tau_j)\mathcal{L}_i(\tau_j, \tau_{j+1})}{\mathcal{L}_i(\tau_{j-1}, \tau_{j+1})}    \right) \right\rbrace,
\end{align}
where $\mathcal{L}_i$ is defined in (\ref{eq:likelihood}). For all $i\in V$ and $j=1, \ldots, k_i+1$, let $\theta_{i, j}=1000$ if $j$ is odd and $\theta_{i, j}=\theta(\mu, k_i)$ if $j$ is even, where $\theta(\mu, k_i)$ is such that $\mu_i=\mu$ for some $\mu>0$ if $i \in C$.  The greater $\mu$, the greater $\theta(\mu, k_i)$. 
A range of values for $\mu$ was considered and Monte Carlo estimations were computed for the corresponding parameters $\theta(\mu, k_i)$. Ten simulations were performed for each combination of the parameters $N, T, v$ and $\mu$.

\begin{figure}[t!]
\centering
\vspace{-4mm}
\hspace{-0.1mm}
\centering
%\hspace{-8mm}
\hspace{10mm}
\begin{tikzpicture}[
Nindex/.style={
circle,
minimum size={0.5cm}
}, 
Caindex/.style={
rectangle, draw, fill = blue!20!white,
minimum size={0.5cm}
}, 
Cbindex/.style={
rectangle, draw, fill = green!20!white,
minimum size={0.5cm}
}, 
Ccindex/.style={
rectangle, draw, fill = red!20!white,
minimum size={0.5cm}
}]
\newcount\rowindex
%row 1
\node[Caindex, draw] (1) at  (1,\rowindex) {};
\node[] at  (1,\rowindex)  {${\scriptstyle 1}$};
\node[Nindex, draw] (7) at  (2,\rowindex)  {};
\node[] at  (2,\rowindex)  {${\scriptstyle 7}$};
\node[Nindex, draw] (13) at  (3,\rowindex)  {};
\node[] at  (3,\rowindex)  {${\scriptstyle 13}$};
\node[Nindex, draw] (19) at  (4,\rowindex)  {};
\node[] at  (4,\rowindex)  {${\scriptstyle 19}$};
\node[Nindex, draw] (25) at  (5,\rowindex)  {};
\node[] at  (5,\rowindex)  {${\scriptstyle 25}$};

%\draw (1) -- (2) -- (3) -- (4) -- (5);

\advance\rowindex by 1
%row 2
\node[Caindex, draw] (2) at  (1,\rowindex) {};
\node[] at  (1,\rowindex)  {${\scriptstyle 2}$};
\node[Nindex, draw] (8) at  (2,\rowindex)  {};
\node[] at  (2,\rowindex)  {${\scriptstyle 8}$};
\node[Ccindex, draw] (14) at  (3,\rowindex)  {};
\node[] at  (3,\rowindex)  {${\scriptstyle 14}$};
\node[Nindex,,draw] (20) at  (4,\rowindex)  {};
\node[] at  (4,\rowindex)  {${\scriptstyle 20}$};
\node[Nindex, draw] (26) at  (5,\rowindex)  {};
\node[] at  (5, \rowindex)  {${\scriptstyle 26}$};

%\draw (1) -- (2) -- (3) -- (4) -- (5);

\advance\rowindex by 1
%row 2
\node[Caindex, draw] (3) at  (1,\rowindex) {};
\node[] at  (1,\rowindex)  {${\scriptstyle 3}$};
\node[Nindex, draw] (9) at  (2,\rowindex)  {};
\node[] at  (2,\rowindex)  {${\scriptstyle 9}$};
\node[Nindex, draw] (15) at  (3,\rowindex)  {};
\node[] at  (3,\rowindex)  {${\scriptstyle 15}$};
\node[Nindex,draw] (21) at  (4,\rowindex)  {};
\node[] at  (4,\rowindex)  {${\scriptstyle 21}$};
\node[Nindex, draw] (27) at  (5,\rowindex)  {};
\node[] at  (5,\rowindex)  {${\scriptstyle 27}$};

\advance\rowindex by 1
%row 2
\node[Caindex, draw] (4) at  (1,\rowindex) {};
\node[] at  (1,\rowindex)  {${\scriptstyle 4}$};
\node[Nindex, draw] (10) at  (2,\rowindex)  {};
\node[] at  (2,\rowindex)  {${\scriptstyle 10}$};
\node[Nindex, draw] (16) at  (3,\rowindex)  {};
\node[] at  (3,\rowindex)  {${\scriptstyle 16}$};
\node[Nindex, draw] (22) at  (4,\rowindex)  {};
\node[] at  (4,\rowindex)  {${\scriptstyle 22}$};
\node[Nindex, draw] (28) at  (5,\rowindex)  {};
\node[] at  (5,\rowindex)  {${\scriptstyle 28}$};

\advance\rowindex by 1
%row 2
\node[Caindex, draw] (5) at  (1,\rowindex) {};
\node[] at  (1,\rowindex)  {${\scriptstyle 5}$};
\node[Nindex, draw] (11) at  (2,\rowindex)  {};
\node[] at  (2,\rowindex)  {${\scriptstyle 11}$};
\node[Cbindex, draw] (17) at  (3,\rowindex)  {};
\node[] at  (3,\rowindex)  {${\scriptstyle 17}$};
\node[Cbindex,,draw] (23) at  (4,\rowindex)  {};
\node[] at  (4,\rowindex)  {${\scriptstyle 23}$};
\node[Cbindex, draw] (29) at  (5,\rowindex)  {};
\node[] at  (5,\rowindex)  {${\scriptstyle 29}$};

\advance\rowindex by 1
%row 2
\node[Caindex, draw] (6) at  (1,\rowindex) {};
\node[] at  (1,\rowindex)  {${\scriptstyle 6}$};
\node[Nindex, draw] (12) at  (2,\rowindex)  {};
\node[] at  (2,\rowindex)  {${\scriptstyle 12}$};
\node[Cbindex, draw] (18) at  (3,\rowindex)  {};
\node[] at  (3,\rowindex)  {${\scriptstyle 18}$};
\node[Cbindex,draw] (24) at  (4,\rowindex)  {};
\node[] at  (4,\rowindex)  {${\scriptstyle 24}$};
\node[Cbindex, draw] (30) at  (5,\rowindex)  {};
\node[] at  (5,\rowindex)  {${\scriptstyle 30}$};

\draw (1) -- (2) -- (3) -- (4) -- (5) -- (6);
\draw (7) -- (8) -- (9) -- (10) -- (11) -- (12);
\draw (13) -- (14) -- (15) -- (16) -- (17) -- (18);
\draw (19) -- (20) -- (21) -- (22) -- (23) -- (24);
\draw (25) -- (26) -- (27) -- (28) -- (29) -- (30);

\draw (6) -- (12) -- (18) -- (24) -- (30);
\draw (5) -- (11) -- (17) -- (23) -- (29);
\draw (4) -- (10) -- (16) -- (22) -- (28);
\draw (3) -- (9) -- (15) -- (21) -- (27);
\draw (2) -- (8) -- (14) -- (20) -- (26);
\draw (1) -- (7) -- (13) -- (19) -- (25);

\node[] at  (3, -0.9)  {(a)};
%\draw (1) -- (2) -- (3) -- (4) -- (5);

%\draw (0,0) circle(7pt) node[]{1} (1,0) circle(7pt) node[]{2} (2,0) circle(7pt) node[]{3} (3,0) circle(7pt) node[]{4} (4,0) circle(7pt) node[]{5};
\end{tikzpicture}
\hspace{20mm}
\begin{tikzpicture}[
Nindex/.style={
circle,
minimum size={0.5cm}
}, 
Caindex/.style={
rectangle, draw, fill = blue!20!white,
minimum size={0.5cm}
}, 
Cbindex/.style={
rectangle, draw, fill = green!20!white,
minimum size={0.5cm}
}, 
Ccindex/.style={
rectangle, draw, fill = red!20!white,
minimum size={0.5cm}
}]
\newcommand{\littleshift}{0.25}

\newcount\rowindex
%row 1
\node[Caindex, draw] (1) at  (1- \littleshift,\rowindex) {};
\node[] at  (1- \littleshift,\rowindex)  {${\scriptstyle 1}$};
\node[Caindex, draw] (3) at  (2- \littleshift,\rowindex)  {};
\node[] at  (2- \littleshift,\rowindex)  {${\scriptstyle 3}$};
\node[Caindex, draw] (5) at  (3- \littleshift,\rowindex)  {};
\node[] at  (3- \littleshift,\rowindex)  {${\scriptstyle 5}$};
\node[Nindex, draw] (7) at  (4- \littleshift,\rowindex)  {};
\node[] at  (4- \littleshift,\rowindex)  {${\scriptstyle 7}$};
\node[Nindex, draw] (9) at  (5- \littleshift,\rowindex)  {};
\node[] at  (5- \littleshift,\rowindex)  {${\scriptstyle 9}$};

\advance\rowindex by 1

%row 1
\node[Caindex, draw] (2) at  (1 + \littleshift,\rowindex) {};
\node[] at  (1+ \littleshift,\rowindex)  {${\scriptstyle 2}$};
\node[Caindex, draw] (4) at  (2+ \littleshift,\rowindex)  {};
\node[] at  (2+ \littleshift,\rowindex)  {${\scriptstyle 4}$};
\node[Caindex, draw] (6) at  (3+ \littleshift,\rowindex)  {};
\node[] at  (3+ \littleshift,\rowindex)  {${\scriptstyle 6}$};
\node[Nindex, draw] (8) at  (4+ \littleshift,\rowindex)  {};
\node[] at  (4+ \littleshift,\rowindex)  {${\scriptstyle 8}$};
\node[Nindex, draw] (10) at  (5+ \littleshift,\rowindex)  {};
\node[] at  (5+ \littleshift,\rowindex)  {${\scriptstyle 10}$};

\advance\rowindex by 1

%row 1
\node[Nindex, draw] (20) at  (1 - \littleshift,\rowindex) {};
\node[] at  (1- \littleshift,\rowindex)  {${\scriptstyle 20}$};
\node[Cbindex, draw] (18) at  (2- \littleshift,\rowindex)  {};
\node[] at  (2- \littleshift,\rowindex)  {${\scriptstyle 18}$};
\node[Nindex, draw] (16) at  (3- \littleshift,\rowindex) {};
\node[] at  (3- \littleshift,\rowindex)  {${\scriptstyle 16}$};
\node[Ccindex, draw] (14) at  (4- \littleshift,\rowindex)  {};
\node[] at  (4- \littleshift,\rowindex)  {${\scriptstyle 14}$};
\node[Nindex, draw] (12) at  (5- \littleshift,\rowindex)  {};
\node[] at  (5- \littleshift,\rowindex)  {${\scriptstyle 12}$};

\advance\rowindex by 1

%row 1
\node[Nindex, draw] (19) at  (1 + \littleshift,\rowindex) {};
\node[] at  (1+ \littleshift,\rowindex)  {${\scriptstyle 19}$};
\node[Cbindex, draw] (17) at  (2+ \littleshift,\rowindex)  {};
\node[] at  (2+ \littleshift,\rowindex)  {${\scriptstyle 17}$};
\node[Nindex, draw] (15) at  (3+ \littleshift,\rowindex) {};
\node[] at  (3+ \littleshift,\rowindex)  {${\scriptstyle 15}$};
\node[Nindex, draw] (13) at  (4+ \littleshift,\rowindex)  {};
\node[] at  (4+ \littleshift,\rowindex)  {${\scriptstyle 13}$};
\node[Nindex, draw] (11) at  (5+ \littleshift,\rowindex)  {};
\node[] at  (5+ \littleshift,\rowindex)  {${\scriptstyle 11}$};

\advance\rowindex by 1

%row 1
\node[Nindex, draw] (21) at  (1 - \littleshift,\rowindex) {};
\node[] at  (1- \littleshift,\rowindex)  {${\scriptstyle 21}$};
\node[Cbindex, draw] (23) at  (2- \littleshift,\rowindex)  {};
\node[] at  (2- \littleshift,\rowindex)  {${\scriptstyle 23}$};
\node[Nindex, draw] (25) at  (3- \littleshift,\rowindex) {};
\node[] at  (3- \littleshift,\rowindex)  {${\scriptstyle 25}$};
\node[Nindex, draw] (27) at  (4- \littleshift,\rowindex)  {};
\node[] at  (4- \littleshift,\rowindex)  {${\scriptstyle 27}$};
\node[Cbindex, draw] (29) at  (5 - \littleshift,\rowindex)  {};
\node[] at  (5 - \littleshift,\rowindex)  {${\scriptstyle 29}$};

% Legend
\node[Caindex] (31) at  (7 - \littleshift,\rowindex)  {};
\node[] at  (7 - \littleshift,\rowindex)  {${i}$};
\node[] at  (7.7 - \littleshift,\rowindex)  {${\in C_1}$};
\node[Cbindex] (32) at  (7 - \littleshift,\rowindex-1)  {};
\node[] at  (7 - \littleshift,\rowindex-1)  {${i}$};
\node[] at  (7.7 - \littleshift, \rowindex-1)  {${\in C_2}$};
\node[Ccindex] (33) at  (7 - \littleshift,\rowindex-2)  {};
\node[] at  (7 - \littleshift,\rowindex-2)  {${i}$};
\node[] at  (7.7 - \littleshift, \rowindex-2)  {${\in C_3}$};
\node[Nindex, draw] (34) at  (7 - \littleshift,\rowindex-3)  {};
\node[] at  (7 - \littleshift,\rowindex-3)  {${i}$};
\node[] at  (7.7 - \littleshift, \rowindex-3)  {${\in \bar{C}}$};

\advance\rowindex by 1

%row 1
\node[Nindex, draw] (22) at  (1 + \littleshift,\rowindex) {};
\node[] at  (1+ \littleshift,\rowindex)  {${\scriptstyle 22}$};
\node[Cbindex, draw] (24) at  (2 + \littleshift,\rowindex)  {};
\node[] at  (2 + \littleshift,\rowindex)  {${\scriptstyle 24}$};
\node[Nindex, draw] (26) at  (3+ \littleshift,\rowindex) {};
\node[] at  (3+ \littleshift,\rowindex)  {${\scriptstyle 26}$};
\node[Nindex, draw] (28) at  (4+ \littleshift,\rowindex)  {};
\node[] at  (4+ \littleshift,\rowindex)  {${\scriptstyle 28}$};
\node[Cbindex, draw] (30) at  (5 + \littleshift,\rowindex)  {};
\node[] at  (5 + \littleshift,\rowindex)  {${\scriptstyle 30}$};

\draw (1) -- (2) -- (3) -- (4) -- (5) -- (6) -- (7) -- (8) -- (9) -- (10);
\draw (10) -- (11) -- (12) -- (13) -- (14) -- (15) -- (16) -- (17) -- (18) -- (19) -- (20);
\draw (20) -- (21) -- (22) -- (23) -- (24) -- (25) -- (26) -- (27) -- (28) -- (29) -- (30);

\draw (1) -- (3) -- (5) -- (7) -- (9) -- (11);
\draw (11) -- (13) -- (15) -- (17) -- (19);
\draw (19) -- (21) -- (23) -- (25) -- (27) -- (29);

\draw (2) -- (4) -- (6)-- (8) -- (10) -- (12) -- (14) -- (16) -- (18) -- (20);
\draw (20) -- (22) -- (24) -- (26)-- (28) -- (30);

\node[] at  (3, -0.9)  {(b)};
\end{tikzpicture}%  
\caption{Graphs of time series indices $i \in \{1, \ldots, 30\}$ representing different dependence structures of changepoints discussed in Section \ref{sec:dependencestructureseg}; panels (a) and (b) correspond to a $6 \times 5$ lattice and a $2$-chain, respectively. For all $i$, circles indicate $i \in \bar{C}$ and squares indicate $i \in C$;  colours blue, green and red indicate that $i\in C_1$, $C_2$ and $C_3$, respectively.}
    \label{fig:graphsimul}
\end{figure}

\subsection{Changepoint inference}
\label{sec:simumcmc}

For each simulation, different models were used to infer changepoint estimates from the data: the proposed changepoint model assuming different prior beliefs on the dependence structure of changepoints, %Inspect \citep{samworth} 
and MVCAPA \citepsupp{fisch}.  
MVCAPA detects collective anomalies in multiple time series such that, by means of lags, anomalies are not necessarily aligned, and it relies on the choice of a penalty parameter that controls the level of evidence required to flag a changepoint. We used the implementation from the \textit{R} package \textit{anomaly} to fit MVCAPA,  without lags and with lags (fixing the maximal lag to 30), with the default penalty rescaled by a constant $\phi$ for various $\phi>0$.

For the proposed changepoint model (\ref{eq:classicconiid}), it is assumed that, for all $i$, $f_i( \, \cdot \, | \,  \theta_{i, j})$ corresponds to $\text{Poisson}(\theta_{i,j})$ and $\theta_{i,j} \sim \Gamma(100, 0.1)$, given the proposed changepoint prior for a collection of changepoint prior parameters. %   for the proposed changepoint prior 
%which encode different prior beliefs on the dependence structure of changepoints, as described below. 
%
Different dependent structures for changepoints across time series were used, as illustrated in Figure \ref{fig:graphsimul}: the dependence structure corresponding to the $r$-chain graph for time series indices, for $r \in \{ 2, 4, 6, \ldots, N/2\}$, given in Section \ref{sec:rchainsegs}; the dependence structure corresponding 
% to a $6 \times (L/6)$ lattice graph for time series indices, 
to a $6 \times 5$ lattice graph, given in Section \ref{sec:latticesegs}, for scenarios with $N=30$ time series; and the dependence structure corresponding to a complete graph given in Section \ref{sec:completegraph}. Moreover, different changepoint prior parameters $\bar{p} \in \{-60, -70, \ldots, -130\}$ and $\lambda =  \lambda_{s}| \bar{p}|/n$ with $\lambda_{s} \in \{0, 0.2, \ldots, 0.8\}$, where $n$ denotes the maximum degree of $G$.  Note that models with $\lambda=0$ correspond to the standard model for independent changepoints (\ref{eq:standard0}). Furthermore, three different scenarios are considered for the upper bounds $\boldsymbol{w}$ for the lags: for the \emph{fixed window} scenario, we fix $w_i = 30$ for all $i$; for the \emph{variable window} scenario, it is assumed \textit{a priori} that $w_i \sim \text{Geometric}(0.9)$ for all $i$; for the \emph{zero window} scenario, we fix $w_i=0$ for all $i$, which is equivalent to assuming the prior for synchronous dependent changepoints defined in (\ref{eq:MRFS}). 

For each combination of changepoint prior parameters, five independent samples of size $2 \, 000 N$ were obtained from the posterior distribution of changepoints, via the MCMC algorithm proposed in Section \ref{sec:binarysampler}, with a burn-in of $300 N$ iterations. Samples were thinned at a rate of one per $N$. %Independent sample chains were compared to assess the convergence of the sampler.
Note that, when $\lambda>0$, we set $\delta_0 = 0.5$, $\delta_1 = 1$, $\delta_2 = 30$ for the prior of the parameter $\delta$, so that linked time series indices are expected to bond with probability $0.5$  when $\delta >0$, according to (\ref{eq:condiaux}).

It is of interest to compare the role of graph-based hyperparameters for the proposed model with the role of $\phi$ for MVCAPA when estimating changepoints. 
%the impact of graph-based parameters on changepoint estimates for the proposed model with the impact of $\phi$ on changepoint estimates for MVCAPA. 
%
MVCAPA and the proposed model provide changepoint estimates for each time series. 
To compare inferred changepoints with respect to the changepoints used to simulate data for each time series, both the mean squared error (MSE)  for the number of changepoints and the loss % function 
$L$ defined in Definition \ref{defn:lossfunction}, fixing $\gamma=40$, which takes into account both the number and the positions of changepoints, are considered. 

\subsection{MCMC diagnostics and runtime}
\label{sec:mcmcdiagnosticruntime}

\begin{figure}[t!]
\vspace{-2mm}
\centering
        \includegraphics[width=0.99\textwidth]{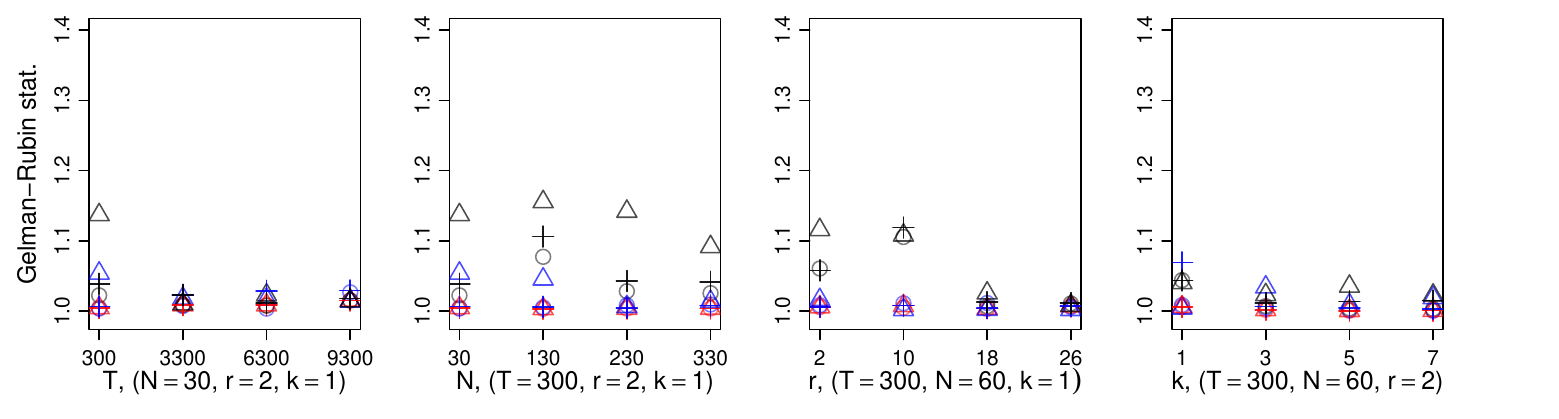}
        \caption{Average Gelman-Rubin statistics for independent samples obtained via the proposed MCMC algorithm for a collection of scenarios: $\lambda_s= 0$ (Red); $\lambda_s =0.2$ (Blue); $\lambda_s =0.6$ (Black); \textit{zero window} (Cross); \textit{fixed window} (Circle); \textit{variable window} (Triangle).}
     \label{fig:gelman}
\end{figure}

\begin{figure}[t!]
\centering
        \includegraphics[width=0.8\textwidth]{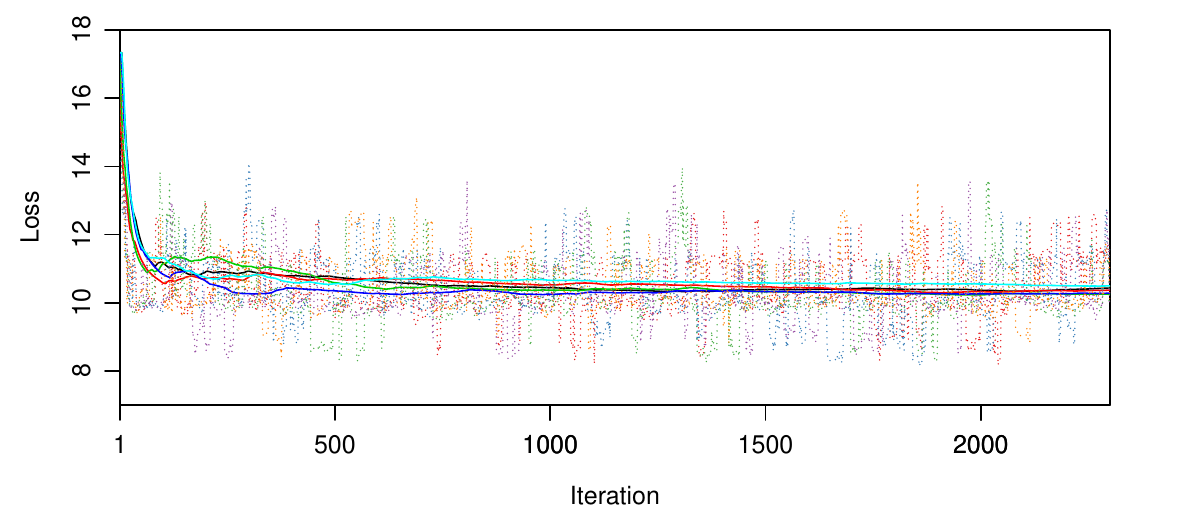}
        \caption{Five independent chains of the value of the loss function obtained via MCMC (dotted lines), with corresponding average loss function value (solid lines). }
     \label{fig:convergenceeg}
\end{figure}

The Gelman-Rubin test \citepsupp{Gelman} is used to assess the convergence of the sampler by comparing between-chain and within-chain variances of multiple chains. For each simulation and choice of changepoint prior parameters, the Gelman and Rubin test statistic, which was computed for the five independent chains of 
 the value of the loss function (\ref{eq:lossfunction})
 obtained via MCMC, was less than $1.2$, suggesting that the sampler has converged; %see Section \ref{sec:appendixmcmc} in the supplementary material \citep{SuppCPGTS} for a breakdown of results. 
 \textcolor{corrections}{see Figure \ref{fig:gelman} for a breakdown of results.} 
Moreover, to illustrate the good mixing properties of the sampler, one simulation for the scenario with $N=30$, $T=300$, $k=1$, $v =10$ and $\mu=90$ is considered. Figure \ref{fig:convergenceeg} displays five independent chains of changepoint loss obtained via the proposed MCMC algorithm assuming the $2$-chain graph for time series with $\bar{p} = -80$, $\lambda_s = 0.6$; % and unknown time window; 
it is apparent that the chains converge to the same changepoints with good mixing properties. 

\begin{figure}[t!]
\centering
        \includegraphics[width=1.05\textwidth]{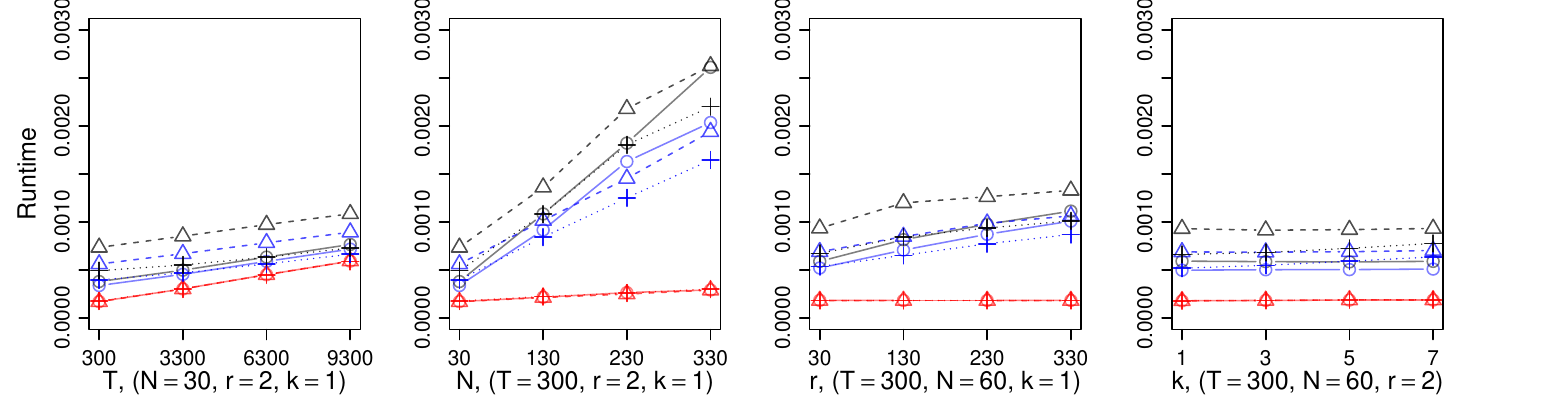}
        \caption{Average runtime in seconds per iteration for the MCMC algorithm for a collection of scenarios: $\lambda_s= 0$ (Red); $\lambda_s =0.2$ (Blue); $\lambda_s =0.6$ (Black); \textit{zero window} (Cross); \textit{fixed window} (Circle); \textit{variable window} (Triangle).}
        %  Red, blue and black for $\lambda=0, 0.2$ and $0.4$, respectively. Cross, circle and triangle for the \textit{zero window}, \textit{fixed window} and \textit{variable window} scenario, respectively.}
     \label{fig:runtimes}
\end{figure}

Figure \ref{fig:runtimes} displays the average runtime of an iteration of the MCMC algorithm, implemented in \textit{Python} 
and run on a 2.6GHz Intel Core i7 processor, for a collection of scenarios assuming an $r$-graph for the dependence structure of the changepoints. Results support the discussion in Section \ref{sec:timecomplexity}. The runtime is linear in the number of observations $T$. When changepoints are dependent with $\lambda >0$ and $\delta >0$, the runtime is linear in the number of time series $N$ and in the number of edges in $G$, which is $2r$ for the $r$-chain graph. However, when $\lambda=0$, we update a single time series at each iteration of the algorithm, so that the runtime is independent of $N$ and $r$. The runtime does not increase with the number of changepoints $k$. Moreover, the runtime per iteration increases for the extended model with lags, for fixed and unknown windows, since lags must be proposed alongside latent changepoints.

%\subsection{Results: detection of clusters of changepoints on a graph}
\subsection{Changepoint estimation results} % detection of clusters of changepoints on a graph}
%By considering dependence structures for changepoints that are induced by the lattice and the $r$-chain graphs
This section discusses results of the simulation study for scenarios with $N=30$ and $T=300$ to demonstrate the merit of the graphical changepoint prior that induces a changepoint model combining weak signals for changepoints across clusters of time series in $G$, in comparison with the standard model for independent changepoints and MVCAPA.

\subsubsection{Detection of clusters of changepoints on a graph of time series} % on a lattice and on an $r$-chain of time series}
\begin{figure}[b!]
\centering
        \includegraphics[width=0.95\textwidth]{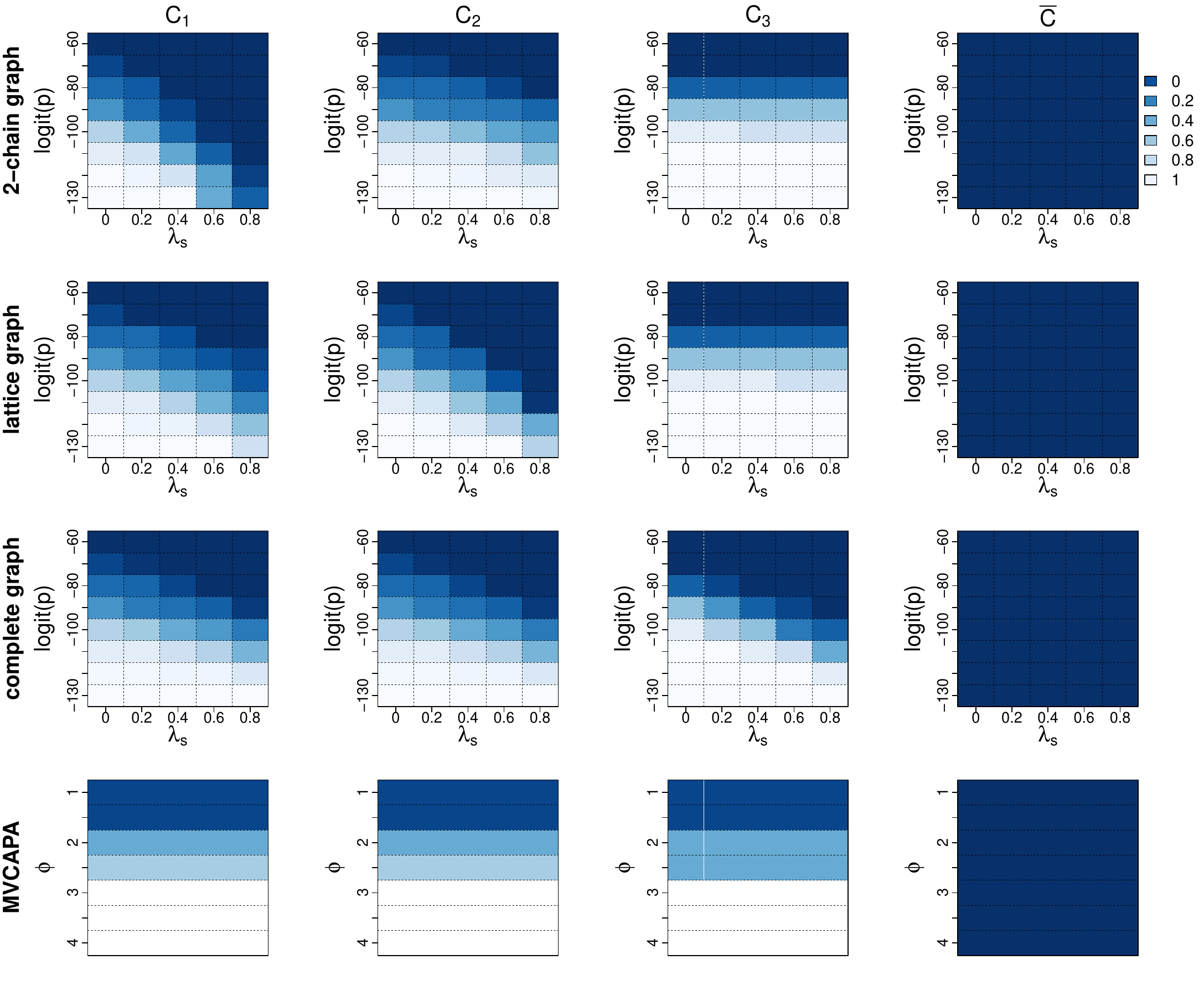}
       \caption{Mean squared error  for the number of changepoints $k_i$ for time series $i$ in clusters $C_1, C_2, C_3, \bar{C}$. Top rows: estimates for the graphical changepoint model assuming the $2$-chain graph, the lattice graph and the complete graph, as a function of $\bar{p} =\text{logit} (p)$ and $\lambda_s$. Bottom row: estimates for MVCAPA as function of $\phi$. Results correspond to scenarios with $v=0$ and $\mu=90$. }
     \label{fig:simucluster}
\end{figure}

\label{sec:experimentsynchclusters}  
First, we focus on demonstrating the role of the graph $G$ and the changepoint parameters $p$ and $\lambda$ using simulations corresponding to scenarios with $v=0$ for synchronous changepoints and with $\mu = 90$.
Figure \ref{fig:simucluster} displays the MSE for $k_i$ for time series $i$ in clusters $C_1, C_2, C_3$, 
for the graphical changepoint models and for MVCAPA as a function of changepoint prior parameters. 

For each graphical changepoint model, as $\bar{p}$ decreases, stronger evidence is required to infer changepoints and therefore the MSE for $k_i$ increases for $i \in  C_1, C_2, C_3$. The impact of $\lambda$ depends on the graph-based dependence structure. Consider the $2$-chain and the $6 \times 5$ lattice graphs for $N=30$ time series as illustrated in Figure \ref{fig:graphsimul}: for $i \in C_3$, the simulated changepoint is isolated on the graphs, whereas for $i \in C_1, C_2$ the simulated changepoints cluster on the graphs. As a result, as $\lambda$ increases, the interaction parameter $\lambda$ has no impact on the MSE for $k_i$ for $i\in C_3$, and the MSE for $k_i$ decreases for $i \in C_1, C_2$ because weaker signals for changepoints, relative to $\bar{p}$, are combined across time series which cluster according to the graphs. The MSE for $k_i$ for $i\in \bar{C}$ is close to $0$ for all scenarios, showing that an increase in  $\lambda$ does not lead to changepoint overfitting. %

However, both for the graphical changepoint model with the complete graph and for MVCAPA, changepoint locations are assumed to be exchangeable across time series,
and therefore the impact of changepoint prior parameters % on the probability that $k_i=1$ 
are identical for all time series $i \in C_1, C_2$ and $C_3$. The simulated changepoints are all connected on the complete graph, and consequently, as $\lambda$ increases, the MSE for $k_i$ decreases for $i \in C_1, C_2, C_3$ for the graphical changepoint model with a complete graph. 
For MVCAPA, as the penalty term $\phi$ increases, the MSE for $k_i$ decreases for time series $i \in C_1, C_2, C_3$. 

%\clearpage

For further evidence that the posterior distribution adapts to the dependence structure for changepoints specified \textit{a priori} via the graph $G$, observe that, as $\lambda$ increases, for the dependence structure induced by the $2$-chain, the MSE for $k_i$ is lower for $i \in C_1$ than for $i \in C_2$, since time series $i\in C_1$ have a greater proportion of neighbour time series impacted by changepoints than time series $i \in C_2$, as illustrated in Figure \ref{fig:graphsimul}. For the dependence structure induced by the lattice, however, time series $i \in C_1$ have a lower proportion of neighbour time series impacted by changepoints than time series $i \in C_2$, so that the MSE for $k_i$ is greater for $i \in C_1$ than for $i \in C_2$. Moreover, as $\lambda$ increases, the MSE for $k_i$ for $i \in C_1$ is lower for the $2$-chain graph than for the complete graph, whereas the MSE for  $k_i$ for $i \in C_2$ is greater for the $2$-chain graph than for the complete graph. %
This follows because the proportion of neighbour time series impacted by changepoints is lower for the complete graph than the $2$-chain graph for time series $i \in C_1$, but larger for time series $i \in C_2$.

\begin{figure}[t!]
\centering
        \includegraphics[width=0.99\textwidth]{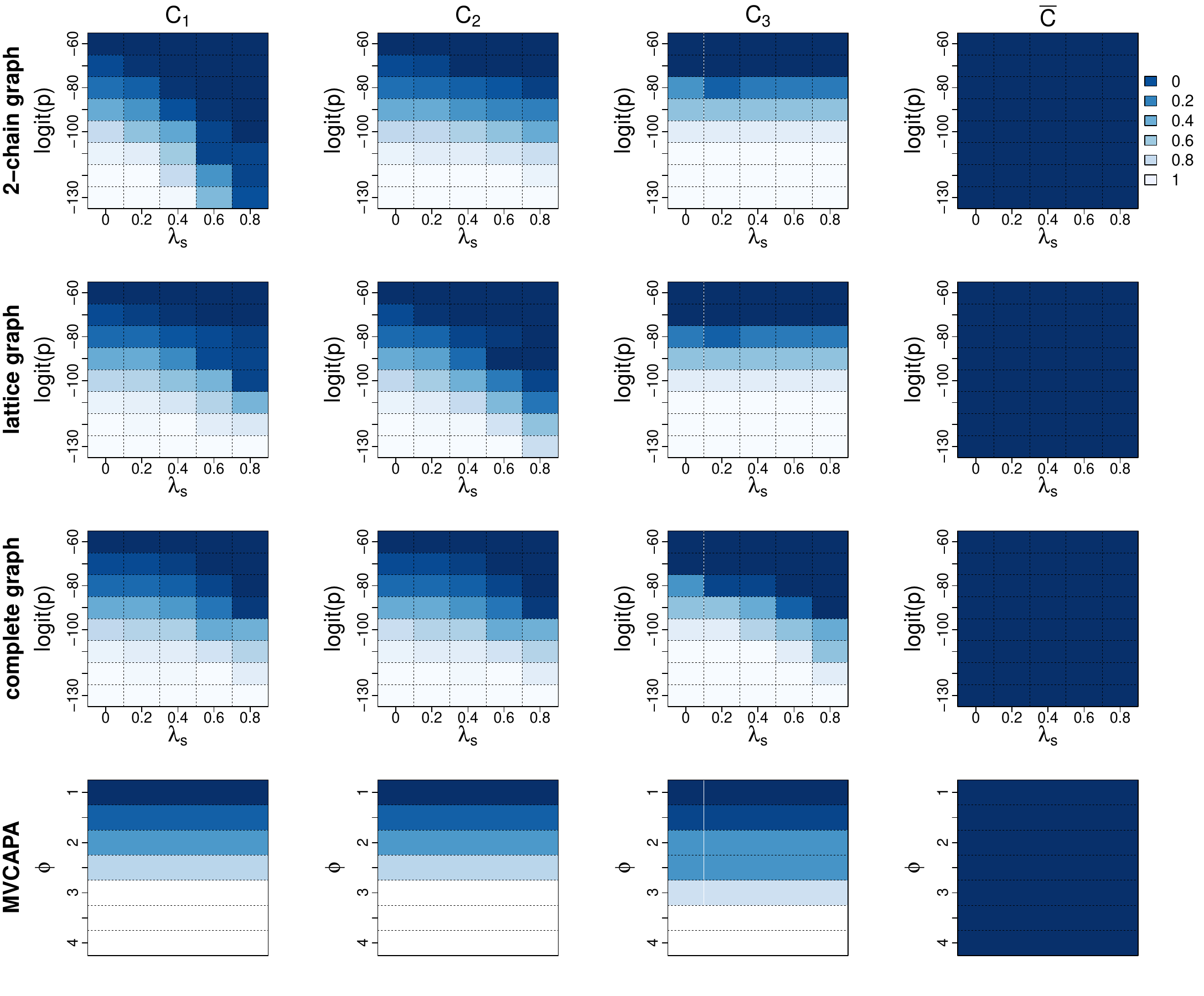}
 \caption{Mean squared error for $k_i$ for time series $i$ in clusters $C_1, C_2, C_3, \bar{C}$. Top rows: estimates for the graphical changepoint model assuming the $2$-chain graph, the lattice graph and the complete graph, as a function of $\bar{p} =\text{logit} (p)$ and $\lambda_s$. Bottom row: estimates for MVCAPA as function of $\phi$. Results correspond to scenarios with $v=10$ and $\mu=90$. }
     \label{fig:simuclusterwithlag}
\end{figure}

The roles of $G$, $p$ and $\lambda$ are the same when detecting asynchronous changepoints. It is apparent in Figure \ref{fig:simuclusterwithlag} , 
that displays the MSE for the number of changepoints  
for the scenarios with $v=10$ for asynchronous changepoints and assuming the \textit{fixed window} scenario for the graphical changepoint models. Results are similar for other levels of asynchrony $v$.

\begin{figure}[h!]
\centering
        \includegraphics[width=0.95\textwidth]{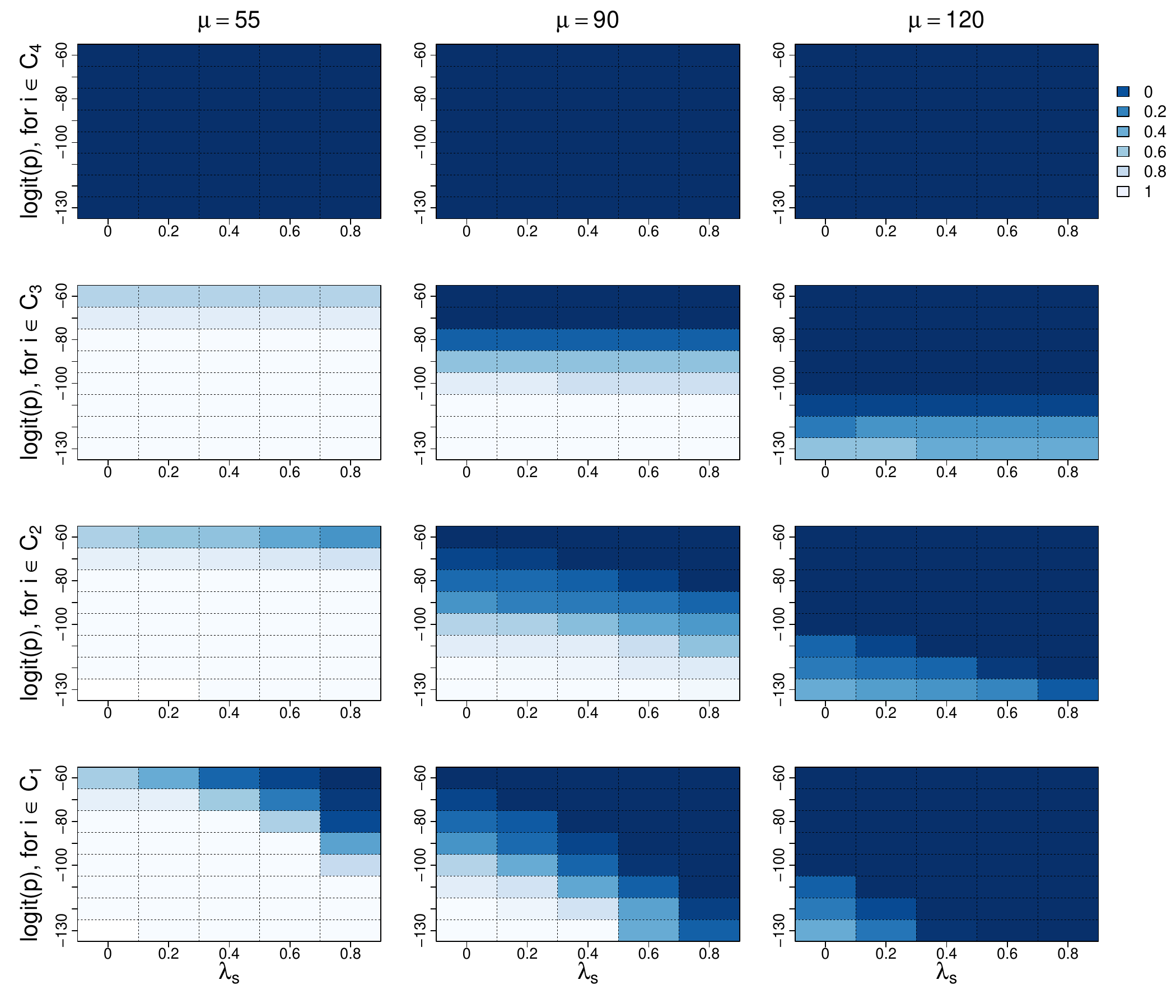}
    \caption{Mean squared error for $k_i$ for times series $i \in C_1, C_3$, given the $2$-chain graph based dependence structure for changepoints, as a function of $\bar{p} =\text{logit} (p)$ and $\lambda_s$,  and for different strength signals for changepoints $\mu \in \{55, 90, 120 \}$.}
     \label{fig:simucluster1}
\end{figure}

For other signal strengths considered in the experiment %discussed in Section \ref{sec:experimentsynchclusters}, corresponding to
 $\mu \in \{ 55, 120\}$, results are similar to results for the scenario where $\mu =90$ when it comes to the role of changepoint prior parameters. Yet, by considering results for the dependence structure corresponding to the $2$-chain displayed in Figure \ref{fig:simucluster1}, we note that, as the signal strength increases, the region of low MSE for $k_i$ for time series $i \in C_1, C_2$ translates along the $\bar{p}$ axis on the $\bar{p},\lambda$-plane.

\clearpage

\begin{figure}[t!]
\centering
        \includegraphics[width=0.95\textwidth]{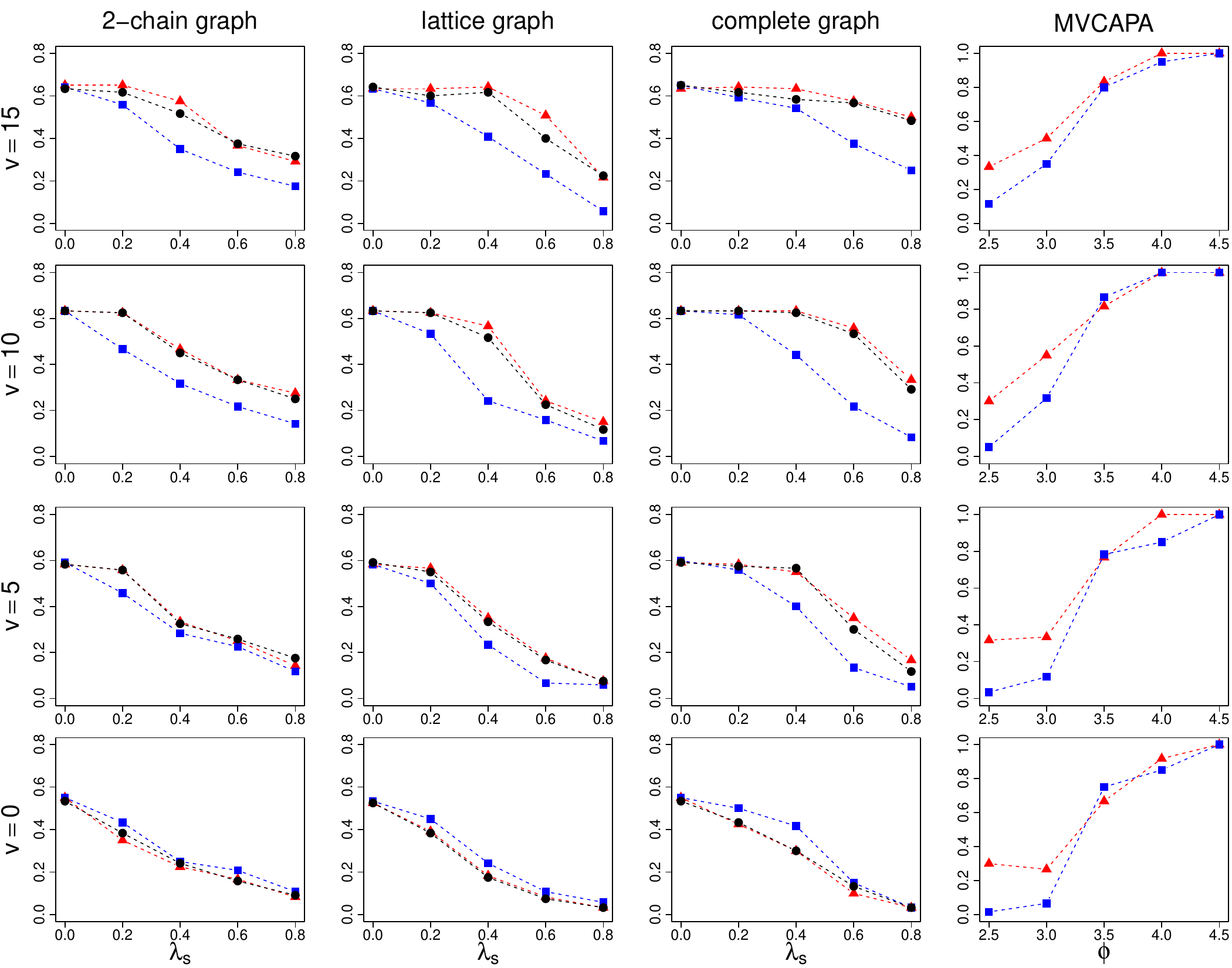}
    %\caption{Monte Carlo estimates of the expected loss with respect to the simulated changepoint parameters for time series $i \in C_1$, given the $2$-chain dependence structure for changepoints, corresponding to different assumptions for the upper bounds for the lags - \emph{fixed window} (blue squares), \emph{variable window} (black circles), and \emph{zero window} (red triangles) - for increasing levels of asynchrony of the changepoints $v$, and for a collection of changepoint hyperparameters $p$, $\lambda_s$. }
     \caption{Mean squared error for $k_i$ %with respect to the simulated changepoint parameters 
  for time series $i \in V$ for increasing levels of changepoint asynchrony $v$. Left columns: graphical changepoint models for different assumptions for the upper bounds for the lags - \emph{fixed window} (blue squares), \emph{variable window} (black circles), and \emph{zero window} (red triangles). Right column:  MVCAPA with lags (blue squares) and without lags (red triangles).}
     \label{fig:simuclusterlag}
     \vspace{-5mm}
\end{figure}

\subsubsection{Detection of quasi-synchronous changepoints}
\label{sec:simuasynchronous}
This section discusses results for scenarios with increasing levels of asynchrony for changepoints $v \in \{0, 5, 10, 15\}$ and with $\mu=90$. Recall changepoints are synchronous for scenarios with $v=0$. 

Figure \ref{fig:simuclusterlag} displays the MSE for $k_i$ for $i \in V$ for the graphical changepoint model assuming different assumptions for the upper bounds for the lags, different graphs $G$ and edge weight parameters $\lambda$, and $\bar{p}=-90$.   
As the level of asynchrony of the changepoints $v$ increases, the MSE for $k_i$ tends to be greater for the \emph{zero window} scenario than for the \emph{fixed window} or the \emph{variable window} scenarios. Hence, 
there is merit in relaxing the assumption that signals for changepoints must be synchronous. In particular, the MSE for $k_i$ tends be greater for the \emph{variable window} than for the \emph{fixed window} scenarios, therefore encouraging practioners to specify fixed time windows when possible.
Moreover, as the interaction parameter $\lambda$ increases, the decrease in MSE tends to be greater for the $2$-chain graph and the lattice graph than for the complete graph. %
This follows because for time series $i \in C$ the proportion of neighbour time series impacted by changepoints is greater for the lattice graph and the $2$-chain graph than for the complete graph.  
Figure \ref{fig:simuclusterlag} also displays results for MVCAPA with and without lags. As the level of asynchrony for changepoints $v$ increases, the MSE tends to increase for MVCAPA without lags. Results also show that, in contrast with the graphical changepoint models, specifying fixed lags when simulated changepoints are synchronous may adversely affect the performance of MVCAPA.

\begin{figure}[t!]
\centering
        \includegraphics[width=0.95\textwidth]{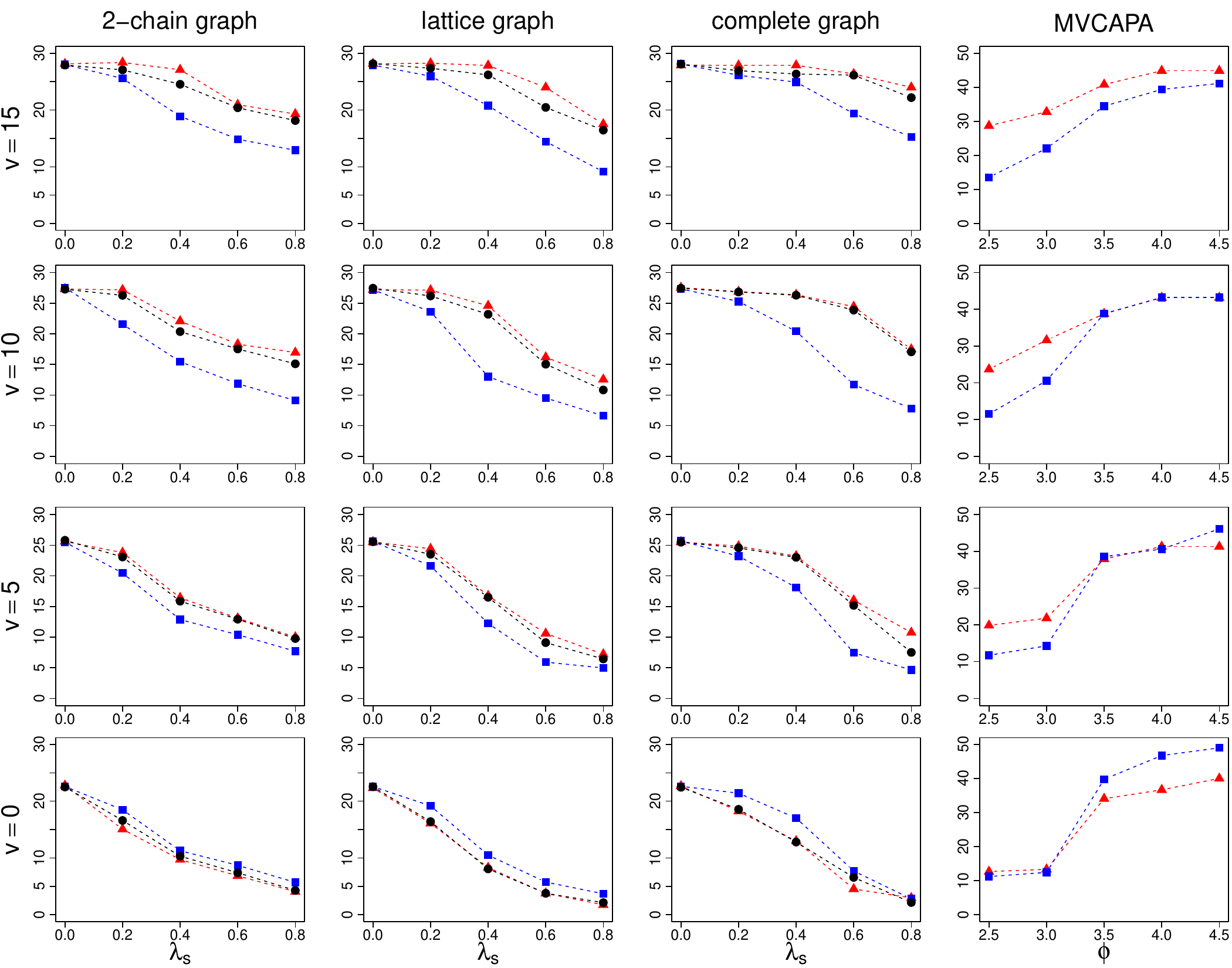}
     \caption{Monte Carlo estimates of the expected loss $L$ 
  for time series $i \in V$ for increasing levels of changepoint asynchrony $v$. Left columns: graphical changepoint models for different assumptions for the upper bounds for the lags - \emph{fixed window} (blue squares), \emph{variable window} (black circles), and \emph{zero window} (red triangles). Right column:  MVCAPA with lags (blue squares) and without lags (red triangles).}
     \label{fig:simuclusterlaglast}
     \vspace{-5mm}
\end{figure}

Figure \ref{fig:simuclusterlaglast} displays the results discussed in Section \ref{sec:simuasynchronous} in terms of the expected loss $L$, showing the model extensions have merit both in terms of detecting asynchronous signals for changepoints and correctly estimating the positions of these signals.

\begin{figure}[t!]
\centering
        \includegraphics[width=0.95\textwidth]{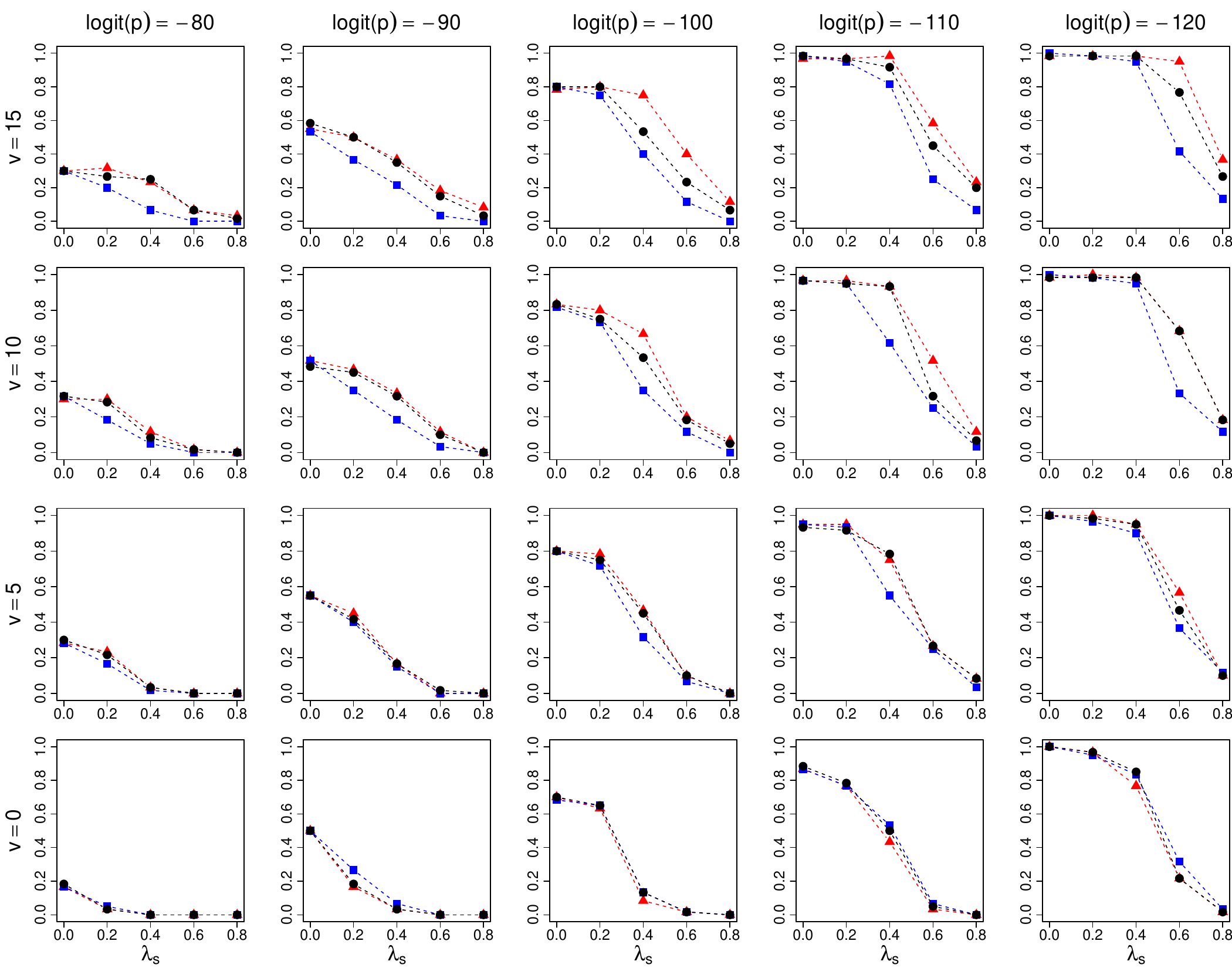}
    \caption{Mean square error for $k_i$  for time series $i \in C_1$, given the $2$-chain dependence structure for changepoints, corresponding to different assumptions for the upper bounds for the lags - \emph{fixed window} (blue squares), \emph{variable window} (black circles), and \emph{zero window} (red triangles) - for increasing levels of asynchrony of the changepoints $v$, and for a collection of changepoint prior parameters $p$, $\lambda_s$.}
     \label{fig:simuclusterlag2}
     \vspace{-5mm}
\end{figure}

 As shown in Figure \ref{fig:simuclusterlag2}, the role of changepoint lags is the same when assuming different $\bar{p}$. %The roles of the graph $G$ and the changepoint parameters $\bar{p}$ and $\lambda$ are discussed   in Section \ref{sec:experimentsynchclusters}. 

\clearpage
\begin{figure}[b!]
\centering
        \includegraphics[width=0.9\textwidth]{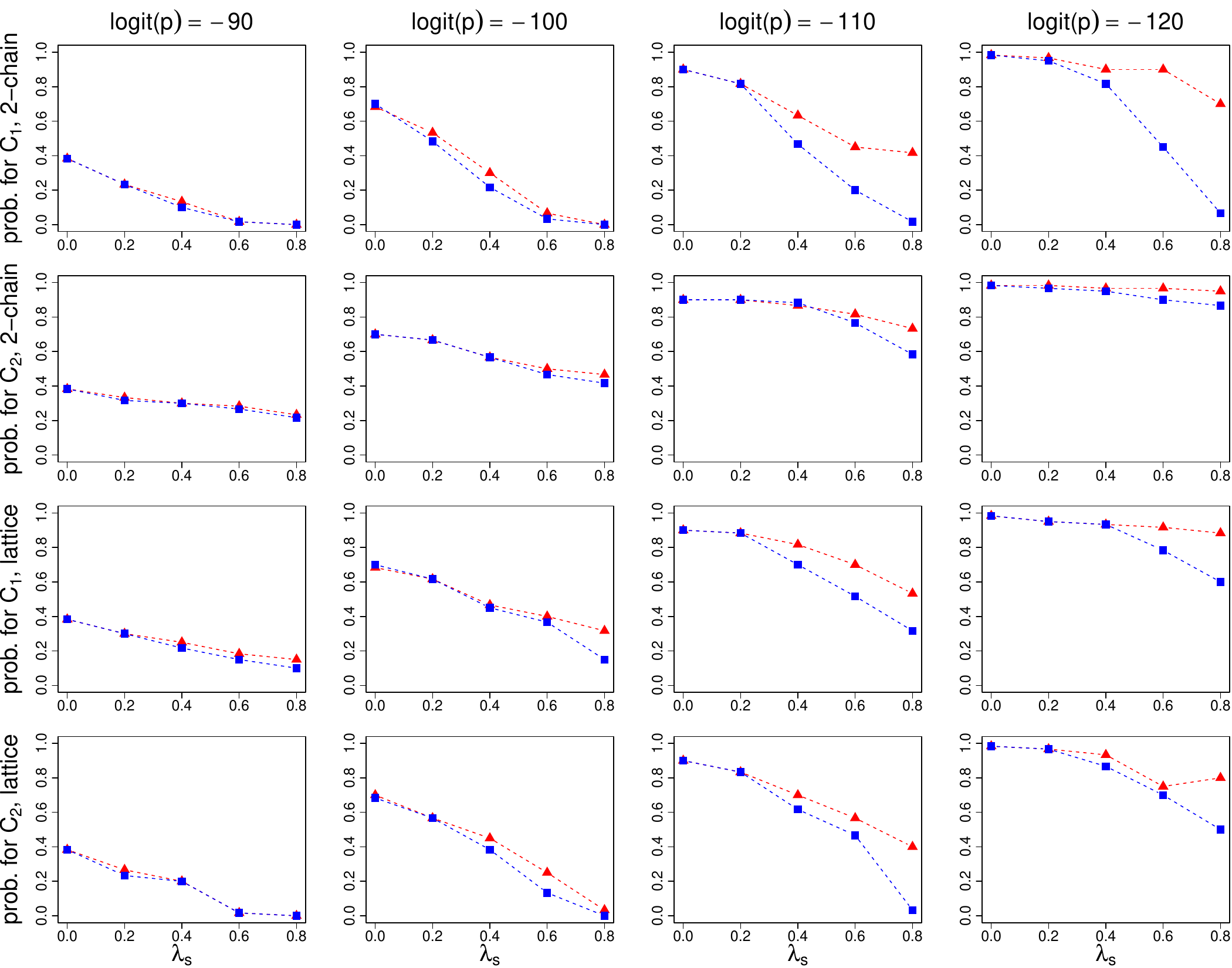}
    \caption{Mean squared error for $k_i$ for time series $i \in C_1, C_2$, obtained via simulations with auxiliary variables (blue squares) and without auxiliary variables (red triangles), for the $2$-chain and the lattice graph based dependence structures, and for a collection of changepoint prior parameters $p$ and $\lambda_s$. }
     \label{fig:simuclusterinfere}
\end{figure}

\vspace{-3mm}

\subsubsection{Importance of auxiliary variables}
\label{sec:simuauxi}
We demonstrate that it is pertinent to use auxiliary variables to sample from the posterior distribution of dependent changepoints, as discussed in Section \ref{sec:inferenceMCMC}. For the simulated data discussed in Section \ref{sec:experimentsynchclusters}, which correspond to scenarios with $v=0$ and $\mu=90$, changepoints were sampled via MCMC as described in Section \ref{sec:simumcmc}, but now with $\delta=0$, meaning the parameter space was not augmented with auxiliary variables (\ref{eq:probq}). %For all samples, the MSE for  $k_i$ for all $i \in \bar{C}$, and that $k_i < 2$ for all $i\in C= C_1 \cup C_2 \cup C_3$ are all greater than $0.99$. 
Figure \ref{fig:simuclusterinfere} compares the MSE for $k_i$ from samples obtained with and without auxiliary variables, for $i \in C_1, C_2$, and for the $2$-chain and the lattice graph based dependence structures for changepoints. 
Without auxiliary variables, the MCMC algorithm fails to explore the state space of  changepoints so that 
%the probabilities that $k_i=1$ 
the number of changepoints 
tend to be underestimated; in other words, clusters of weak signals for changepoints across time series tend to be overlooked. The difference in performance, in favour of the MCMC algorithm making use of auxiliary variables, is particularly important when interaction parameters are large and weak signals for changepoints correspond to time series whose indices induce subgraphs with a large number of edges, as for $i \in C_1$ given the $2$-chain dependence structure in Figure \ref{fig:simuclusterinfere}.

% and the Heilbronn Institute for Mathematical Research.
%The authors acknowledge funding from the Los Alamos National Laboratory, EPSRC and the Heilbronn Institute for Mathematical Research. %Research presented in this article was supported by the Laboratory Directed Research and Development program of Los Alamos National Laboratory (New Mexico, USA) under project number 20180607ECR.
\clearpage
%\bibliographystyle{ba}
%\bibliography{bibli.bib}
%\bibliographystylesupp{ba}
\bibliographysupp{bibli.bib}

\end{document}